\documentclass[twoside]{IEEEtran}
\usepackage{units}
\usepackage{bm}
\usepackage{amsmath}
\usepackage{amssymb}
\usepackage{graphicx}
\usepackage{esint}
\usepackage{mathrsfs}
\usepackage{cite}

\makeatletter

\providecommand{\tabularnewline}{\\}
   
\newtheorem{thm}{\protect\theoremname}
\newtheorem{rem}{\protect\remarkname}
\newtheorem{cor}{\protect\corollaryname}
\newtheorem{prop}{\protect\propositionname}
\newtheorem{lem}{\protect\lemmaname}

\makeatother

\providecommand{\corollaryname}{Corollary}
\providecommand{\lemmaname}{Lemma}
\providecommand{\propositionname}{Proposition}
\providecommand{\remarkname}{Remark}
\providecommand{\theoremname}{Theorem}

\begin{document}

\title{Outage Exponent: A Unified Performance Metric\\for Parallel Fading
Channels}

\author{Bo (Bob) Bai,~\IEEEmembership{Member,~IEEE,}
        Wei Chen,~\IEEEmembership{Member,~IEEE,}\\
        Khaled B. Letaief,~\IEEEmembership{Fellow,~IEEE,} and
        Zhigang Cao,~\IEEEmembership{Seiner~Member,~IEEE}%
\thanks{Bo (Bob) Bai, Wei Chen and Zhigang Cao are with Tsinghua National Laboratory for Information Science and Technology (TNList), and Department of Electronic Engineering, Tsinghua University, Beijing 100084, China. E-mail: eebobai@tsinghua.edu.cn, wchen@tsinghua.edu.cn, czg-dee@tsinghua.edu.cn.

Khaled B. Letaief is with Center for Wireless Information Technology, and Department of Electronic and Computer
Engineering, Hong Kong University of Science and Technology (HKUST), Clear Water Bay, Kowloon, Hong Kong. E-mail: eekhaled@ust.hk.

Bo (Bob) Bai is also with HKUST when this work was finished.

This work was supported partly by NSFC key projects under Grant No.
60832008, Grant No. 60902001, and Grant No. 61021001, and RGC project under Grant No. 610311.

This paper was presented in part at IEEE ICC 2010 and IEEE Globecom 2010.

Manuscript received September 12, 2011; revised May 21, 2012.

Copyright (c) 2012 IEEE. Personal use of this material is permitted.%
}}
\maketitle
\begin{abstract}
The parallel fading channel, which consists of finite number of subchannels,
is very important, because it can be used to formulate many practical
communication systems. The outage probability, on the other hand,
is widely used to analyze the relationship among the communication
efficiency, reliability, SNR, and channel fading. To the best of our
knowledge, the previous works only studied the asymptotic outage performance
of the parallel fading channel which are only valid for a large number
of subchannels or high SNRs. In this paper, a unified performance
metric, which we shall refer to as the outage exponent, will be proposed.
Our approach is mainly based on the large deviations theory and the Meijer's
$G$-function. It is shown that the proposed outage exponent is not
only an accurate estimation of the outage probability for any number
of subchannels, any SNR, and any target transmission rate, but also
provides an easy way to compute the outage capacity, finite-SNR diversity-multiplexing
tradeoff, and SNR gain. The asymptotic performance metrics, such as
the delay-limited capacity, ergodic capacity, and diversity-multiplexing
tradeoff can be directly obtained by letting the number of subchannels
or SNR tends to infinity. Similar to Gallager's error exponent, a
reliable function for parallel fading channels, which illustrates
a fundamental relationship between the transmission reliability and
efficiency, can also be defined from the outage exponent. Therefore,
the proposed outage exponent provides a complete and comprehensive
performance measure for parallel fading channels.\end{abstract}

\markboth{IEEE Transactions on Information Theory, vol. XX, no. XX, Month YEAR}{Outage
Exponent: A Unified Performance Metric for the Parallel Fading Channel}

\section{Introduction}

\PARstart{T}{he} parallel fading channel has a finite number of flat
fading subchannels, where the channel gain of each subchannel only
depends on its own fading statistics. In \cite{Kaplan1995}, this
model is also referred to as the block-fading channel. The parallel
fading channel is very important because many practical communication
systems can be formulated into this model. The conventional narrow-band
system, such as GSM, can be modeled as a parallel fading channel in
the time domain \cite{Ozarow1994}. The wide-band OFDM system, such
as WiMAX and 3GPP LTE, is a parallel fading channel in the frequency
domain \cite{Wong1999}. By applying a singular value decomposition,
the MIMO channel can also be formulated into a parallel fading channel
in the space domain \cite{Tse2005}.

For the parallel fading channel, the ergodic capacity cannot be achieved
because it only has a finite number of subchannels which yields a
non-ergodic fading case \cite{Biglieri1998}. Therefore, the outage
probability, defined as the probability that the instantaneous channel
capacity is smaller than a target transmission rate, becomes a fundamental
performance metric for the non-ergodic parallel fading channel \cite{Ozarow1994}.
From the outage probability perspective, many important performance
parameters, such as outage capacity, delay-limited capacity (or zero-outage
capacity), ergodic capacity, diversity-multiplexing tradeoff, and
finite-SNR diversity-multiplexing tradeoff, can be obtained directly.
However, it is very difficult to accurately calculate the outage probability
for the parallel fading channel except for two trivial cases, i.e.,
only one subchannel, and infinity number of subchannels. In \cite{Ozarow1994},
Ozarow, Shamai, and Wyner gave an integration formula for the outage
probability of the parallel fading channel with two subchannels. This
result is the first and can be considered as a milestone step. From
then on, many works have been published on this topic. The results
can be roughly divided into two categories: 1) the outage probability
versus the number of subchannels; and 2) the outage performance versus
the signal-to-noise ratio (SNR).

When considering the outage probability for any number of subchannels,
Kaplan and Shamai provided an upper bound from the Chernoff bounding
method in \cite{Kaplan1995}. As the authors noticed, however, the
upper bound is not tight. Another problem of this result lies in the
fact that the target transmission rate should deviate from the ergodic
capacity largely. Inspired by the idea in \cite{Gallager1965}, some
works also tried to estimate the error probability for the parallel
fading channel from the theory of error exponent. The original result
for the ergodic parallel channel can be found in \cite{Gallager1968}.
Divsalar and Biglieri then proposed upper bounds on the error probability
of coded systems over AWGN and fading channels in \cite{Divsalar2000}.
A similar work can also be found in \cite{Lun2002}. Based on the
second type of the Duman-Salehi (DS-2) bound in \cite{Duman1998}, Sason
and Shamai proposed improved bounds on the decoding error probability
of block codes over fully-interleaved fading channels \cite{Sason2001}.
They also evaluated the proposed bounds on turbo-like and LDPC codes
in \cite{Sason2001}. By applying a similar approach, Wu, Xiang, and
Ling proposed a new upper bound for block-fading channels in \cite{Wu2007},
which is tight for the channel with a large number of subchannels,
i.e., the near-ergodic case. An excellent survey of this approach
can be found in \cite{Shamai2002}. Because the theory of error exponent
considers both the coding scheme and channel fading at the same time,
the proposed bounds are often very complicated and not tight enough.
Hence, it is not trivial to provide insights for the parallel fading
channel clearly and directly from these results.

Another approach is to study the outage performance versus SNR, in
which the ideal coding scheme is assumed to be used. An important
result is to evaluate the diversity-multiplexing tradeoff for fading
channels, where each point of the tradeoff curve is just the slope
of the outage probability for a given multiplexing gain (or normalized
target rate) as SNR tends to infinity. This concept was first proposed
for MIMO channels \cite{Zheng2003}, and the corresponding results
for the parallel fading channel can be found in \cite{Tse2005}. Since
the diversity-multiplexing tradeoff is valid in the high SNR regime,
the corresponding finite-SNR version for MIMO channels is independently
proposed in \cite{Narasimhan2006} and \cite{Loyka2007}, respectively.
They have a similar definition and can converge to the diversity-multiplexing
tradeoff when SNR tends to infinity. The finite-SNR diversity-multiplexing
tradeoff can be used to estimate the additional SNR required to decrease
the outage probability by a specified amount for a given multiplexing
gain. This approach does not estimate the coefficient of the exponential
function, which means it cannot be used to estimate the SNR gain for
different coding schemes when SNR is not high enough.

In this paper, a unified performance metric for parallel fading channels,
which we shall refer to as the outage exponent, will be proposed in
order to analyze the relationship among the outage performance, the
number of subchannels, and SNR at the same time. The proposed outage
exponent has many advantages. It only focuses on the fading effect
of the channel, and hence it is much tighter and simpler than the
error exponent approach. The outage exponent also provides an accurate
estimation of the outage probability for any number of subchannels,
any SNR, and any target transmission rate. Similar to the error exponent,
a reliable function for the parallel fading channel can then be defined
to illustrate the fundamental relationship between the communication
efficiency and reliability. From this reliable function, we will show
that: 1) the outage probability will tend to zero as the number of
subchannels tends to infinity, if and only if the average target rate
is smaller than the ergodic capacity; and 2) the outage probability
will tend to zero as the SNR tends to infinity for any average target
rate lower than the capacity of additive white Gaussian noise (AWGN)
channels. Furthermore, the outage capacity, finite-SNR diversity-multiplexing
tradeoff, and SNR gain can also be obtained from the outage exponent.
Then, the asymptotic performance metrics, such as the delay-limited
capacity, ergodic capacity, and diversity-multiplexing tradeoff are
just the limits of the previous results, and can be directly obtained
by letting the number of subchannels or SNR tend to infinity. Therefore,
the proposed outage exponent provides a complete performance framework
for parallel fading channels. It also provides a powerful tool for
analyzing and evaluating the performance of existing and upcoming
communication systems.

In order to analyze the outage exponent for the parallel fading channel,
we must consider two different cases. First of all, the outage exponent
is analyzed for the case where the target rate is smaller than the
ergodic capacity. Inspired by the successful application of large
deviations theory on analyzing the bit-error probability for avalanche
photodiode receivers \cite{Letaief1992}, the latest results of large
deviations theory in \cite{Bucklew1993} and \cite{Theodosopoulos2007}
are used to calculate tight upper and lower bounds on the outage probability,
respectively. For the case where the target rate is close to and greater
than the ergodic capacity, the Meijer's $G$-function and the method
of integral around a contour in \cite{Gradshteyn2007} are used to
compute the upper and lower bounds. In order to achieve the above
calculated outage exponent, the coding schemes which maximize the
minimum product distance are also discussed. The proposed method combines
the advantages of the rotated $\mathbb{Z}^{L}$-lattices code and
permutation code \cite{Oggier2004,Tavildar2006}.

The rest of the paper is organized as follows. Section \ref{sec:model and problem}
presents the system model and the precise problem formulation. In
Section \ref{sec:outage exponent}, the general definition and related
properties of the outage exponent are presented. Section \ref{sec:smaller case}
studies the outage exponent when the target rate is smaller than the
ergodic capacity. The results of delay-limited capacity, ergodic capacity,
and diversity-multiplexing tradeoff are also presented in this section.
In Section \ref{sec:larger case}, the outage exponent and the reliable
function are studied when the target rate is higher than the ergodic
capacity. Section \ref{sec:error exponent} will illustrate the differences
between the proposed outage exponent and the error exponent. Section
\ref{sec:coding} studies some coding issues in the parallel fading
channel. In Section \ref{sec:numerical}, numerical results are provided
to verify the theoretical derivations. Finally, Section \ref{sec:conclusions}
concludes the paper.

\section{System Model and Problem Formulation\label{sec:model and problem}}

\subsection{Parallel Fading Channel Model}

Consider a parallel fading channel with $L$ subchannels, each of
which undergoes independent flat Rayleigh fading. In narrowband systems,
each subchannel may correspond to the duration of coherence time.
In broadband systems, each subchannel corresponds to one coherence
bandwidth in the slow fading scenario, or one coherence bandwidth
in the duration of coherence time in the block-fading scenario. For
convenience, we assume that each subchannel has a unit time duration
and a unit bandwidth throughout this paper. In addition, we assume
that the perfect channel state information (CSI) is only known at
the receiver side.

Let $h_{l},\, l=1,\ldots,L$ denote the channel gain of the $l$th
subchannel. Then, $h_{l}$ is a random variable with the circularly
symmetric Gaussian distribution $\mathcal{CN}\left(0,1\right)$, and
$h_{l}$ is independent with $h_{l'}$ if $l\neq l'$. The row vector
$\bm{x}_{l},\, l=1,2,\ldots,L$ denotes the transmission symbols over
the $l$th subchannel, while the row vector $\bm{y}_{l}$ denotes
the corresponding received symbols. The parallel fading channel can
then be modeled by

\begin{equation}
\bm{Y}=\bm{H}\bm{X}+\bm{W},\label{eq:block-fading channel}
\end{equation}
where
\[
\bm{Y}=\left(\begin{array}{c}
\bm{y}_{1}\\
\bm{y}_{2}\\
\vdots\\
\bm{y}_{L}
\end{array}\right),\qquad\bm{X}=\left(\begin{array}{c}
\bm{x}_{1}\\
\bm{x}_{2}\\
\vdots\\
\bm{x}_{L}
\end{array}\right),
\]

\[
\bm{H}=\mathrm{diag}\left(h_{1},h_{2},\ldots,h_{L}\right).
\]
$\bm{W}$ is the white Gaussian noise matrix, where the elements of
$\bm{W}$ are independent with the identical distribution of $\mathcal{CN}\left(0,1\right)$.
Hence, the mutual information between the transmitter and the receiver,
denoted by $I\left(\bm{H}\right)$, is then given by
\[
I\left(\bm{H}\right)=\sum_{l=1}^{L}\ln\left(1+\left|h_{l}\right|^{2}\gamma\right),
\]
where $\gamma$ is the received SNR. Throughout this paper, the natural
logarithmic function $\ln\left(x\right)$ is used, and the unit of
information is ``$\mathrm{nat}$''.

\subsection{Outage Formulation}

The outage probability is an important concept in fading channels,
which provides a way to characterize the performance of communication
systems in non-ergodic fading scenarios. Clearly, the parallel fading
channel is non-ergodic when $L$ is finite. According to \cite{Biglieri1998},
the outage probability of the parallel fading channel is defined by

\begin{equation}
\begin{aligned}p_{\mathrm{out}}\left(L,\gamma,R\right) & =\Pr\left\{ I\left(\bm{H}\right)<R\right\} \\
 & =\Pr\left\{ \sum_{l=1}^{L}\ln\left(1+\left|h_{l}\right|^{2}\gamma\right)<R\right\} \\
 & =\Pr\left\{ \frac{1}{L}\sum_{l=1}^{L}\ln\left(1+\left|h_{l}\right|^{2}\gamma\right)<\overline{R}\right\} ,
\end{aligned}
\label{eq:outage definition}
\end{equation}
where $R$ is the target transmission rate or coding rate, and $\overline{R}=\frac{R}{L}$
is the average rate on each subchannel. For convenience, $p_{\mathrm{out}}$
will be used in the following instead of $p_{\mathrm{out}}\left(L,\gamma,R\right)$.

This definition characterizes the relationship among the outage probability
$p_{\mathrm{out}}$, the transmission rate $R$, the number of subchannels
$L$, and the SNR $\gamma$. The outage probability and the transmission
rate represent two key performance metrics for communication systems,
i.e., the reliability and the efficiency, respectively. The number
of subchannels determines the time interval and the bandwidth used
by the transmission signal, i.e., the degree of freedom. At the same
time, it also determines how many independent channel gains the transmission
signal may undergo, i.e., the diversity order. The SNR represents
the effective energy contained in the signal. Therefore, the outage
formulation contains the fundamental elements which govern the transmission
reliability and efficiency of non-ergodic fading channels.

Unfortunately, it is very difficult to derive the exact formula for
the outage probability as defined in Eq. \eqref{eq:outage definition}.
By now, only some approximations have been proposed in previous works.
In \cite{Kaplan1995}, from the Chernoff's bound, an upper bound of
the outage probability is given by
\begin{equation}
p_{\mathrm{out}}<\min_{\lambda\geq0}\left\{ e^{\lambda\overline{R}}\left[\gamma^{-\frac{\lambda}{2L}}e^{\frac{1}{2\gamma}}W_{-\frac{\lambda}{2L},\frac{L-\lambda}{2L}}\left(\gamma^{-1}\right)\right]^{L}\right\} ,\label{eq:outage bound Shamai}
\end{equation}
where $W_{\nu,\mu}\left(z\right)$ is the Whittaker's function \cite{Gradshteyn2007}.
As the authors noticed in \cite{Kaplan1995}, however, the bound in
Eq. \eqref{eq:outage bound Shamai} is not tight. In \cite{Tse2005},
a lower bound is given by
\begin{equation}
\begin{aligned}p_{\mathrm{out}} & >\left(\Pr\left\{ \ln\left(1+\left|h_{l}\right|^{2}\gamma\right)<\overline{R}\right\} \right)^{L}\\
 & =\left(1-e^{-\frac{e^{\overline{R}}-1}{\gamma}}\right)^{L}.
\end{aligned}
\label{eq:outage bound Tse}
\end{equation}
However, Eq. \eqref{eq:outage bound Tse} is only an approximation
of the outage probability when SNR tends to infinity, which is not
tight in realistic SNRs.

\section{Outage Exponent\label{sec:outage exponent}}

As stated before, it is very difficult to compute the exact outage
probability for the parallel fading channel directly. Likewise, it
is also very tough to accurately analyze the decoding error probability
for a given coding scheme. To overcome the difficulties, Gallager
proposed a systematical approach to estimate the upper and lower bounds
for the decoding error probability, which is often referred to as
the \emph{error exponent} \cite{Gallager1968}. Similarly, this paper
tries to propose an \emph{outage exponent} approach to calculate the
exponentially tight upper and lower bounds for the outage probability
in non-ergodic fading channels.

\subsection{General Results}

The general result on the outage exponent is given below in Theorem \ref{thm:outage exponent}.
For convenience, we let the symbol
``$\lesssim$'' denote the following relationship:
\begin{equation}
f\left(x\right)\lesssim g\left(x\right)\Leftrightarrow\begin{cases}
f\left(x\right)\leq g\left(x\right);\\
\lim_{x\rightarrow\infty}\frac{f\left(x\right)}{g\left(x\right)}=1.
\end{cases}\label{eq:leq_app}
\end{equation}
Similarly, we also define the symbol ``$\gtrsim$''.
\begin{thm}
\label{thm:outage exponent}For a parallel fading channel with $L$
subchannels, the outage exponents are the exponentially tight upper
and lower bounds of the outage probability
\begin{equation}
\left\{ \begin{aligned} & p_{\mathrm{out}}\lesssim p_{\mathrm{ex}}^{\mathrm{upper}}=\psi e^{-L\left[E_{1}\left(R,\gamma\right)+\frac{E_{0}\left(\gamma\right)}{L}+o\left(L\right)\right]},\\
 & p_{\mathrm{out}}\gtrsim p_{\mathrm{ex}}^{\mathrm{lower}}=\varphi e^{-L\left[E_{1}\left(R,\gamma\right)+\frac{E_{0}\left(\gamma\right)}{L}+o\left(L\right)\right]};
\end{aligned}
\right.\label{eq:outage exponent}
\end{equation}
where $\varphi$ and $\psi$ are constants or slowly varying functions
with $\varphi\leq\psi$. $p_{\mathrm{ex}}^{\mathrm{upper}}$ and $p_{\mathrm{ex}}^{\mathrm{lower}}$
are referred to as the upper and lower outage exponents, respectively.
The function $E_{1}\left(R,\gamma\right)$ is given by
\begin{equation}
E_{1}\left(R,\gamma\right)=\left(C_{\mathrm{awgn}}-\overline{R}\right)E_{1,1}\left(\gamma\right)+E_{1,0}\left(\gamma\right),
\end{equation}
where $E_{1}\left(R,\gamma\right)>0$ if and only if $\overline{R}<\overline{C}$
with $\overline{C}$ being the ergodic capacity of each subchannel,
and $E_{1,1}\left(\gamma\right)$ and $E_{1,0}\left(\gamma\right)$
satisfy
\begin{equation}
\left\{ \begin{alignedat}{1} & \lim_{\gamma\rightarrow\infty}\frac{C_{\mathrm{awgn}}E_{1,1}\left(\gamma\right)}{\ln\gamma}=1;\\
 & \lim_{\gamma\rightarrow\infty}\frac{E_{1,0}\left(\gamma\right)}{\ln\gamma}=0.
\end{alignedat}
\right.
\end{equation}
The function \textup{$E_{0}\left(\gamma\right)$ satisfies
\begin{equation}
\lim_{\gamma\rightarrow\infty}\frac{E_{0}\left(\gamma\right)}{\ln\gamma}=0.
\end{equation}
}Finally, the function $o\left(L\right)$ tends to zero when $L\rightarrow\infty$.
\end{thm}

Theorem \ref{thm:outage exponent} captures the intrinsic principles
of the outage probability from two dimensions, namely, the number of subchannels
$L$, and the SNR $\gamma$.

\subsection{Relationship with Other Performance Metrics}

The proposed outage exponent integrates many important performance
metrics. As such, it gives a complete picture of the comprehensive performance
for parallel fading channels. Based on Theorem \ref{thm:outage exponent},
the relationship between the outage exponent and other performance
metrics are discussed in this subsection.

\subsubsection{Outage Capacity}

For a given outage probability $\varepsilon$, the outage capacity
is defined as the supremum of the transmission rate that satisfies
$p_{\mathrm{out}}<\varepsilon$ in \cite{Biglieri1998}. Therefore,
the $\varepsilon$-outage capacity, denoted by $C_{\varepsilon}$,
is given by
\begin{equation}
C_{\varepsilon}=\sup\left\{ \overline{R}:p_{\mathrm{out}}<\varepsilon\right\} ,\label{eq:outage capacity}
\end{equation}
where $\sup\mbox{\ensuremath{\mathcal{A}}}$ is the supremum of the
set $\mathcal{A}$. To obtain the outage capacity, an accurate estimation of the outage probability
is needed. Note that this is given by the outage exponent in Theorem \ref{thm:outage exponent}.

\subsubsection{Delay-Limited Capacity (or Zero-Outage Capacity)}

In \cite{Biglieri1998}, the delay-limited capacity, denoted by $C_{\mathrm{dl}}$,
which is also known as the zero-outage capacity, is the maximum transmission
rate as the outage probability tends to zero when $L\rightarrow\infty$.
Therefore, $C_{\mathrm{dl}}$ is defined as 
\[
C_{\mathrm{dl}}=\sup\left\{ \overline{R}:\lim_{L\rightarrow\infty}p_{\mathrm{out}}=0\right\} ,
\]
Clearly, $C_{\mathrm{dl}}=\lim_{\varepsilon\rightarrow0}\lim_{L\rightarrow\infty}C_{\varepsilon}$.
By applying the error exponent in Theorem \ref{tab:outage exponent},
we have 
\[
\lim_{L\rightarrow\infty}\frac{1}{L}\ln p_{\mathrm{out}}=E_{1}\left(R,\gamma\right).
\]
Hence, the delay-limited capacity can be obtained as
\begin{equation}
C_{\mathrm{dl}}=\sup\left\{ \overline{R}:E_{1}\left(R,\gamma\right)>0\right\} ,\label{eq:delay limited capacity}
\end{equation}

\subsubsection{Ergodic Capacity}

In \cite{Biglieri1998}, the ergodic capacity for fading channels,
denoted by $\overline{C}$, is defined as the statistical average
of the channel capacity. According to the central limit theorem, we
have 
\[
\overline{C}=\lim_{L\rightarrow\infty}\frac{I\left(\bm{H}\right)}{L}=\mathsf{E}\left\{ \ln\left(1+\left|h_{l}\right|^{2}\gamma\right)\right\} .
\]
From Theorem \ref{thm:outage exponent}, if $\overline{R}<\overline{C}$,
then $p_{\mathrm{out}}\rightarrow0$ as $L\rightarrow\infty$, which
implies that $C_{\mathrm{dl}}\geq\overline{C}$. On the other hand,
if $E_{1}\left(R,\gamma\right)>0$, we have $\overline{R}<\overline{C}$,
which implies that $C_{\mathrm{dl}}\leq\overline{C}$. As a result,
the delay limited capacity is equal to the ergodic capacity in parallel
fading channels, that is 
\begin{equation}
C_{\mathrm{dl}}=\overline{C}=\mathsf{E}\left\{ \ln\left(1+\left|h_{l}\right|^{2}\gamma\right)\right\} .\label{eq:ergodic capacity}
\end{equation}

\subsubsection{Reliable Function}

\begin{figure}[t]
\centering

\includegraphics[width=3.5in]{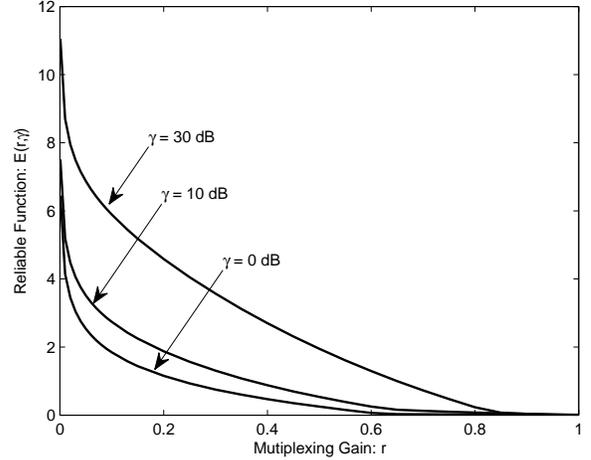}

\caption{Reliable function for the parallel fading channel with $\gamma=0,10,30\,\mathrm{dB}$,
and $L=4$.\label{fig:reliable}}
\end{figure}

In \cite{Gallager1968}, Gallager proposed an error exponent to reveal
the fundamental principle that the decoding error probability exponentially
varies with the coding length. The outage exponent, however, studies
the exponentially decreasing of the outage probability as the number
of fading subchannels increases. Therefore, the outage exponent reveals
the fundamental principle that the outage probability varies with
the ergodicity of the channel, i.e., from slow fading, block fading,
to fast fading. Similar to the error exponent, the \emph{reliable
function} for the outage exponent can be defined as follows
\begin{equation}
E\left(R,\gamma\right)=E_{1}\left(R,\gamma\right)+\frac{E_{0}\left(\gamma\right)}{L}+o\left(L\right).\label{eq:reliable func}
\end{equation}
If we let
\[
r=\frac{R}{C_{\mathrm{awgn}}},
\]
where $r$ is referred to as the multiplexing gain, the reliable function
can also be denoted by $E\left(r,\gamma\right)$. From the properties
of $E\left(R,\gamma\right)$ in Theorem \ref{thm:outage exponent},
we have the following results.
\begin{itemize}
\item For any fixed SNR, if $\overline{R}$ is smaller than $\overline{C}$,
the outage probability will tend to zero as $L\rightarrow\infty$.
\item For any fixed $L$, if $\overline{R}$ is smaller than $C_{\mathrm{awgn}}$,
the outage probability will tend to zero as $\gamma\rightarrow\infty$. 
\end{itemize}

An example of the reliable function versus multiplexing gain is plotted
in Fig. \ref{fig:reliable} for different SNRs. It can be seen that
for any given $r$, the reliable function gets larger as the SNR increases.
But all of the curves tend to zero as $r\rightarrow1$, which is the
same as the error exponent figures in \cite{Gallager1968}.

\subsubsection{Finite-SNR Diversity-Multiplexing Tradeoff}

In realistic SNRs, the finite-SNR diversity-multiplexing tradeoff,
proposed by \cite{Narasimhan2006,Loyka2007} for MIMO channels, is
used to estimate the additional SNR required to decrease the outage
probability by a specified amount for a given multiplexing gain. In
\cite{Narasimhan2006}, the finite-SNR diversity-multiplexing tradeoff
is defined as
\[
d_{\mathrm{f}}^{*}\left(r\right)=-\frac{\partial\ln p_{\mathrm{out}}}{\partial\ln\gamma},
\]
which is derived by estimating the upper bound of the outage probability.
From Theorem \ref{thm:outage exponent}, the finite-SNR diversity-multiplexing
tradeoff for parallel fading channels can be obtained by
\begin{equation}
d_{\mathrm{f}}^{*}\left(r\right)=L\gamma\frac{\partial E\left(r,\gamma\right)}{\partial\gamma},\label{eq:finite-SNR dmt def}
\end{equation}
where $d_{\mathrm{f}}^{*}\left(r\right)$ represents
the reliable function in Eq. \eqref{eq:reliable func} from another
perspective.

\subsubsection{Diversity-Multiplexing Tradeoff}

In \cite{Zheng2003}, the diversity-multiplexing tradeoff is defined
as the slope of the outage probability when SNR tends to infinity.
This concept reveals the fundamental relationship between reliability
and efficiency in the asymptotic case. From \cite{Tse2005}, the optimal
diversity-multiplexing tradeoff for parallel fading channels is given
by

\begin{equation}
d^{*}\left(r\right)=-\lim_{\gamma\rightarrow\infty}\frac{\ln p_{\mathrm{out}}}{\ln\gamma}=L\left(1-\frac{r}{L}\right).\label{eq:dmt def}
\end{equation}
This result was obtained by applying the lower bound in Eq. \eqref{eq:outage bound Tse}.
From Theorem \ref{thm:outage exponent}, $d^{*}\left(r\right)$ can
be derived as follows
\[
\begin{aligned}d^{*}\left(r\right) & =\lim_{\gamma\rightarrow\infty}-\frac{\ln p_{\mathrm{out}}}{\ln\gamma}=\lim_{\gamma\rightarrow\infty}-\frac{LE\left(R,\gamma\right)}{\ln\gamma}\\
 & =\lim_{\gamma\rightarrow\infty}\frac{L\left[E_{1}\left(R,\gamma\right)+\frac{E_{0}\left(\gamma\right)}{L}+o\left(L\right)\right]}{\ln\gamma}\\
 & =\lim_{\gamma\rightarrow\infty}\frac{L\left[\left(1-\frac{r}{L}\right)E_{1,1}\left(\gamma\right)\ln\left(1+\gamma\right)+E_{1,0}\left(\gamma\right)\right]}{\ln\gamma}\\
 & =L\left(1-\frac{r}{L}\right).
\end{aligned}
\]
By applying the properties of the outage exponent, it is not difficult
to verify that 
\[
d^{*}\left(r\right)=\lim_{\gamma\rightarrow\infty}d_{\mathrm{f}}^{*}\left(r\right)=\lim_{\gamma\rightarrow\infty}L\gamma\frac{\partial E\left(r,\gamma\right)}{\partial\gamma}.
\]

\subsubsection{SNR Gain (or Coding Gain)}

The SNR gain is the difference between the SNR values needed by two
different coding schemes to achieve a given outage probability. This
metric is very useful to evaluate coding schemes with the same diversity
gain. In the high SNR regime, the coding gain is determined by the
coefficient of the outage exponent.

\section{Outage Exponent for $\overline{R}<\overline{C}$\label{sec:smaller case}}

The outage exponent for the parallel fading channel will be studied
thoroughly in Sections \ref{sec:smaller case} and \ref{sec:larger case}.
According to our study, the channel outage behavior is different for
$\overline{R}<\overline{C}$ and $\overline{R}\geq\overline{C}$.
Thus, the outage exponent will be analyzed in two different cases.
For the $\overline{R}<\overline{C}$ case, our derivation is mainly
based on the latest results in large deviations theory and Meijer's
$G$-function \cite{Dembo1998,Gradshteyn2007}.

Before presenting the main results, we introduce some basic terminologies
and notations for this section. Let $\left\{ Y_{n}:n\in\mathbb{N}\right\} $
be a sequence of real-valued random variables with distribution $F_{n}\left(y\right)$.
The logarithmic moment generating function of $Y_{n}$ is given by
\[
\Lambda_{n}\left(\xi\right)=\ln M_{n}\left(\xi\right)=\ln\mathsf{E}\left\{ e^{\xi Y_{n}}\right\} .
\]
The Legendre-Fenchel transform is then defined as
\[
\Lambda_{n}^{*}\left(s\right)=\sup_{\xi\in\mathbb{R}}\left\{ \xi s-\Lambda_{n}\left(\xi\right)\right\} .
\]
Define the Legendre duality as
\[
\Xi\left(s\right)=\underset{\xi\in\mathbb{R}}{\arg\sup}\left\{ \xi s-\Lambda_{n}\left(\xi\right)\right\} .
\]
The limit of the logarithmic moment generating function is given by
\[
\Lambda\left(\xi\right)=\lim_{n\rightarrow\infty}\frac{1}{n}\Lambda_{n}\left(\xi\right).
\]
Finally, define the titled distribution of $Y_{n}$ as

\begin{equation}
dF_{n}^{\left(\xi\right)}\left(y\right)=\frac{e^{\xi y}dF_{n}\left(y\right)}{M_{n}\left(\xi\right)},\label{eq:titled}
\end{equation}
and let $Y_{n}^{\left(\xi\right)}$ be a random variable having $F_{n}^{\left(\xi\right)}\left(y\right)$
as its distribution.

\subsection{Upper Outage Exponent}

The upper outage exponent for the parallel fading channel, defined
as the exponentially tight upper bound of the outage probability,
is given in the following theorem.
\begin{thm}
\label{thm:upper bound 1}For any $R$ with $\overline{R}<\overline{C}$,
the upper outage exponent $p_{\mathrm{ex}}^{\mathrm{upper}}$ for
a parallel fading channel with $L$ subchannels is given by

\begin{equation}
p_{\mathrm{out}}\lesssim p_{\mathrm{ex}}^{\mathrm{upper}}=\frac{1}{\sqrt{2\pi}}e^{-L\left[E_{1}\left(R,\gamma\right)+\frac{\ln\left(\sigma\Xi\left(0\right)\right)}{L}+\frac{\ln L}{2L}\right]},\label{eq:upper bound 1}
\end{equation}
where
\begin{equation}
\begin{aligned}E_{1}\left(R,\gamma\right)= & \left(C_{\mathrm{awgn}}-\overline{R}\right)\Xi\left(0\right)-\Xi\left(0\right)\ln\left(1+\frac{1}{\gamma}\right)\\
 & -\ln\left(e^{\frac{1}{\gamma}}\Gamma\left(1-\Xi\left(0\right),\gamma^{-1}\right)\right).
\end{aligned}
\label{eq:reliable function}
\end{equation}
The parameter $\Xi\left(0\right)$ is the unique positive solution
of the following equation
\begin{equation}
\overline{R}-\frac{1}{\Gamma\left(1-\xi,\gamma^{-1}\right)}G_{2,3}^{3,0}\left(\frac{1}{\gamma}\left|\begin{aligned} & 1,1\\
 & 0,0,1-\xi
\end{aligned}
\right.\right)=0,\label{eq:Xi0}
\end{equation}
while $\sigma^{2}$ is given by
\begin{equation}
\begin{aligned}\sigma^{2}= & \frac{2}{\Gamma\left(1-\Xi\left(0\right),\gamma^{-1}\right)}G_{3,4}^{4,0}\left(\frac{1}{\gamma}\left|\begin{aligned} & 1,1,1\\
 & 0,0,0,1-\Xi\left(0\right)
\end{aligned}
\right.\right)-\\
 & \left[\frac{1}{\Gamma\left(1-\Xi\left(0\right),\gamma^{-1}\right)}G_{2,3}^{3,0}\left(\frac{1}{\gamma}\left|\begin{aligned} & 1,1\\
 & 0,0,1-\Xi\left(0\right)
\end{aligned}
\right.\right)\right]^{2}.
\end{aligned}
\label{eq:sigma}
\end{equation}
In these equations, $\Gamma\left(z,\alpha\right)$ is the incomplete
Gamma function, which is defined by 
\[
\Gamma\left(z,\alpha\right)=\int_{\alpha}^{\infty}e^{-t}t^{z-1}dt,
\]
and
\[
\begin{aligned} & G_{p,q}^{m,n}\left(z\left|\begin{aligned}a_{1}, & \ldots,a_{p}\\
b_{1}, & \ldots,b_{q}
\end{aligned}
\right.\right)=\\
 & \frac{1}{2\pi i}\oint_{\mathcal{L}}\frac{\prod_{j=1}^{m}\Gamma\left(b_{j}-s\right)\prod_{j=1}^{n}\Gamma\left(1-a_{j}+s\right)}{\prod_{j=m+1}^{q}\Gamma\left(1-b_{j}+s\right)\prod_{j=n+1}^{p}\Gamma\left(a_{j}-s\right)}z^{s}ds,
\end{aligned}
\]
is the Meijer's $G$-function.\end{thm}
\begin{IEEEproof}
See Appendix \ref{app:upper bound 1}.\end{IEEEproof}
\begin{rem}
The upper outage exponent proposed in Theorem
\ref{thm:upper bound 1} is also known as the second order saddle-point
approximation for any given $L$ and $\gamma$. The validity and accuracy
of this approximation for finite $L$ and $\gamma$ is justified in
\cite{Butler2007}.
\end{rem}

By solving the equation $\varepsilon=p_{\mathrm{out}}\lesssim p_{\mathrm{ex}}^{\mathrm{upper}}$,
the following corollary can be obtained immediately.
\begin{cor}
For a parallel fading channel with $L$ subchannels, the $\varepsilon$-outage
capacity is given by
\begin{equation}
C_{\varepsilon}\gtrsim\ln\frac{\gamma}{\left[\left(\varepsilon\sqrt{2\pi L}\sigma\Xi\left(0\right)\right)^{\frac{1}{L}}e^{\frac{1}{\gamma}}\Gamma\left(1-\Xi\left(0\right),\gamma^{-1}\right)\right]^{\frac{1}{\Xi\left(0\right)}}}.\label{eq:capacity}
\end{equation}

\end{cor}

According to Theorem \ref{thm:upper bound 1}, other two corollaries
which illustrate the asymptotic performance of the parallel fading
channel from two different dimensions can also be obtained directly.
\begin{cor}
For a parallel fading channel with $L$ subchannels, the delay-limited
capacity is equal to the ergodic capacity when $L\rightarrow\infty$,
i.e.,
\begin{equation}
C_{\mathrm{dl}}=\overline{C}=\mathsf{E}\left\{ \ln\left(1+\left|h_{l}\right|^{2}\gamma\right)\right\} .\label{eq:dl ergodic}
\end{equation}
\end{cor}
\begin{IEEEproof}
From Theorem \ref{thm:upper bound 1}, if $\overline{R}<\overline{C}$,
then $p_{\mathrm{out}}\rightarrow0$ as $L\rightarrow\infty$. Therefore,
we have $C_{\mathrm{dl}}\geq\overline{C}$.

On the other hand, we have $E_{1}\left(R,\gamma\right)=\Lambda^{*}\left(0\right)$.
According to the definition of Legendre-Fenchel transform, $\Lambda^{*}\left(s\right)$
is non-deceasing for $s>\mathsf{E}\left\{ Y_{L}\right\} =\overline{R}-\overline{C}$.
Therefore, if $E_{1}\left(R,\gamma\right)>0$, we have $\overline{R}<\overline{C}$,
i.e., $C_{\mathrm{dl}}\leq\overline{C}$. As a result, Eq. \eqref{eq:dl ergodic}
holds.\end{IEEEproof}
\begin{cor}
The diversity-multiplexing tradeoff for a parallel fading channel
with $L$ subchannels is given by
\begin{equation}
d^{*}\left(r\right)=L\left(1-\frac{r}{L}\right).\label{eq:dmt}
\end{equation}
\end{cor}
\begin{IEEEproof}
According to the definition of diversity-multiplexing tradeoff and
Theorem \ref{thm:upper bound 1}, we have
\[
d^{*}\left(r\right)=\lim_{\gamma\rightarrow\infty}\frac{LE\left(R,\gamma\right)}{\ln\gamma}.
\]
From Eq. \eqref{eq:reliable func}, it is not difficult to verify
the following equations:
\[
\lim_{\gamma\rightarrow\infty}\frac{C_{\mathrm{awgn}}\Xi\left(0\right)}{\ln\gamma}=\lim_{\gamma\rightarrow\infty}\frac{\ln\left(1+\gamma\right)\Xi\left(0\right)}{\ln\gamma}=1,
\]

\[
\lim_{\gamma\rightarrow\infty}\frac{\Xi\left(0\right)\ln\left(1+\frac{1}{\gamma}\right)}{\ln\gamma}=0,
\]
\[
\lim_{\gamma\rightarrow\infty}\frac{\ln\left(e^{\frac{1}{\gamma}}\Gamma\left(1-\Xi\left(0\right),\gamma^{-1}\right)\right)}{\ln\gamma}=0,
\]
\[
\lim_{\gamma\rightarrow\infty}\frac{\frac{\ln\left(\sigma\Xi\left(0\right)\right)}{L}+\frac{\ln L}{2L}}{\ln\gamma}=0.
\]
Therefore, Eq. \eqref{eq:dmt def} holds.
\end{IEEEproof}

\subsection{Lower Outage Exponent\label{sub:lower bound}}

As stated before, the outage exponent also depends on the exponentially
tight lower bound. According to the proof in Appendix \ref{app:upper bound 1},
we have 
\[
\varphi\lesssim\frac{p_{\mathrm{out}}}{e^{-L\left[E_{1}\left(R,\gamma\right)+\frac{\ln\left(\sigma\Xi\left(0\right)\right)}{L}+\frac{\ln L}{2L}\right]}}\lesssim\frac{1}{\sqrt{2\pi}}.
\]
Thus, we only need to give a good estimation on $\varphi$ for the
lower outage exponent. The results are summarized in the following
theorem.
\begin{thm}
\label{thm:lower bound 1}For any $R$ with $\overline{R}<\overline{C}$,
the lower outage exponent $p_{\mathrm{ex}}^{\mathrm{lower}}$ for
a parallel fading channel with $L$ subchannels is given by

\begin{equation}
\begin{aligned} & p_{\mathrm{out}}\gtrsim p_{\mathrm{ex}}^{\mathrm{lower}}\\
 & =\sup_{0<\alpha<1}\left\{ \left(1-e^{-\Lambda_{\alpha}^{*}\left(\delta R\right)}-e^{-\Lambda_{\alpha}^{*}\left(R\right)}\right)e^{-LE_{1}^{\alpha}\left(R,\gamma\right)}\right\} ,
\end{aligned}
\label{eq:lower bound 1}
\end{equation}
where
\begin{equation}
\delta=\frac{\alpha-e^{-\Lambda_{\alpha}^{*}\left(R\right)}}{1-e^{-\Lambda_{\alpha}^{*}\left(R\right)}},\label{eq:delta}
\end{equation}
and
\[
\left\{ \begin{aligned}E_{1}^{\alpha}\left(R,\gamma\right)= & \left(C_{\mathrm{awgn}}-\delta\overline{R}\right)\Xi\left(\alpha R\right)-\Xi\left(\alpha R\right)\ln\left(1+\frac{1}{\gamma}\right)\\
 & +\ln\left(e^{\frac{1}{\gamma}}\Gamma\left(1+\Xi\left(\alpha R\right),\gamma^{-1}\right)\right);\\
\Lambda_{\alpha}^{*}\left(\delta R\right)= & R\left(\alpha-\delta\right)\Xi\left(\alpha R\right)-\int_{\delta R}^{\alpha R}\Xi\left(t\right)dt;\\
\Lambda_{\alpha}^{*}\left(R\right)= & \left.\Lambda_{\alpha}^{*}\left(\delta R\right)\right|_{\delta=1}.
\end{aligned}
\right.
\]
The parameter $\Xi\left(t\right)$ is the negative solution of the
following equation
\[
t-\frac{L}{\Gamma\left(1+\xi,\gamma^{-1}\right)}G_{2,3}^{3,0}\left(\frac{1}{\gamma}\left|\begin{aligned} & 1,1\\
 & 0,0,1+\xi
\end{aligned}
\right.\right)=0.
\]
\end{thm}
\begin{IEEEproof}
See Appendix \ref{app:lower bound 1}.
\end{IEEEproof}

From the proof of Theorem \ref{thm:lower bound 1}, it can be seen
that $\Xi\left(R\right)$ here is equal to $-\Xi\left(0\right)$ in
Theorem \ref{thm:upper bound 1}, which implies that $E_{1}^{\alpha}\left(R,\gamma\right)$
is the same as $E_{1}\left(R,\gamma\right)$. Therefore, the lower
bound and the upper bound of the outage probability have the same
exponent. The only difference lies in the coefficient of the main
exponential function. A key step in the proof, i.e., Eq. \eqref{eq:chernoff},
can also be replaced by applying the same technique in the proof of
Theorem \ref{thm:upper bound 1}, which will result in a tighter bound.

\section{Outage Exponent for $\overline{R}\geq\overline{C}$\label{sec:larger case}}

It is well known that if the coding rate is larger than the ergodic
capacity, the decoding error probability of communication over ergodic
channels is still greater than a constant when the coding length tends
to infinity. For the parallel fading channel, the behavior of the
outage probability when the target transmission rate is higher than
the ergodic capacity is still unknown.

\subsection{Upper and Lower Outage Exponents}

If the large deviation theory is applied for $\overline{R}>\overline{C}$,
the bounds will be obtained in the form of non outage probability,
i.e., the dualities of Eqns. \eqref{eq:upper bound 1} and \eqref{eq:lower bound 1}.
These results will show that the outage probability tends to one when
$L\rightarrow\infty$ in the $\overline{R}>\overline{C}$ case. Furthermore,
the large deviation bounds will be loose when $\overline{R}$ is close
to $\overline{C}$. Especially, the large deviation bound will be
one, if $\overline{R}=\overline{C}$. Therefore, we need to find another
method to study the behavior of the outage probability when SNR changes.
The results are summarized in the following theorem.
\begin{thm}
\label{thm:Meijer's G bounds}For any $R$ with $\overline{R}\geq\overline{C}$,
the outage probability of a parallel fading channel with $L$ subchannels
can be bounded by
\begin{equation}
\left\{ \begin{aligned} & p_{\mathrm{out}}\lesssim p_{\mathrm{ex}}^{\mathrm{upper}}=1-F_{L}\left(\frac{e^{R}-1}{\gamma^{L}}\right),\\
 & p_{\mathrm{out}}\gtrsim p_{\mathrm{ex}}^{\mathrm{lower}}=1-e^{\frac{L}{\gamma}}F_{L}\left(\frac{e^{R}}{\gamma^{L}}\right),
\end{aligned}
\right.\label{eq:Meijer's G bounds}
\end{equation}
for the high SNR regime. In these equations, $F_{L}\left(z\right)$
is given by 
\[
F_{L}\left(z\right)=G_{0,L}^{L,0}\left(z\left|\begin{array}{c}
-\\
0,1,\ldots,1
\end{array}\right.\right).
\]
In the low SNR regime, the outage probability can be approximated
by
\begin{equation}
p_{\mathrm{out}}\approx1-\frac{\Gamma\left(L,r\right)}{\left(L-1\right)!},\label{eq:Meijer's G_lower}
\end{equation}
where $r=\frac{R}{C_{\mathrm{awgn}}}$ is the multiplexing gain.\end{thm}
\begin{IEEEproof}
See Appendix \ref{app:Meijer's G bounds}.
\end{IEEEproof}

It can be seen that the low SNR approximation is irrelevant to $\gamma$.
In the high SNR regime, the difference between the lower bound and
the upper bound is only the coefficient of the Meijer's $G$-function.
The coefficient in the lower bound is $e^{\frac{L}{\gamma}}$, while
the coefficient in the upper bound is $e^{\frac{0}{\gamma}}=1$. Since
$e^{\frac{L}{\gamma}}\rightarrow1$ as $\gamma\rightarrow\infty$,
the lower and upper bounds will converge to the true value in the
high SNR regime. However, in the low SNR regime, $e^{\frac{0}{\gamma}}$
is too small, while $e^{\frac{L}{\gamma}}$ is too large. Therefore,
if we can modify the coefficient to a proper value, a more accurate
approximation can then be obtained for the low SNR regime.
\begin{prop}
\label{pro:Meijer's G approximation}For any $R$ with $\overline{R}\geq\overline{C}$,
the outage probability of a parallel fading channel with $L$ subchannels
can be approximated by
\begin{equation}
p_{\mathrm{out}}\approx1-e^{\frac{\lambda}{\gamma}}G_{0,L}^{L,0}\left(\frac{e^{R}}{\gamma^{L}}\left|\begin{array}{c}
-\\
0,1,\ldots,1
\end{array}\right.\right),\label{eq:Meijer's G approximation}
\end{equation}
in the high SNR regime, where $0<\lambda<L$.
\end{prop}

For the parallel fading channel with $L$ subchannels, it is very
difficult to estimate $\lambda$ in theory. However, if we set the
simulation value of the outage probability in the low SNR regime equal to $0\,\mathrm{dB}$ in (\ref{eq:Meijer's G approximation}), an estimation of $\lambda$ can be obtained.

\subsection{Reliable Function}

The reliable function for the $\overline{R}\geq\overline{C}$ case
will be analyzed in this subsection. Since the outage probability
tends to one when $L\rightarrow\infty$ in this case, we will mainly
focus on the behavior of the reliable function versus SNR. Before
we study the reliable function, a basic lemma will be given first.
\begin{lem}
\label{lem:convex}Let $f\left(x\right)$ be a strict concave function
in $\left[0,\infty\right)$. If $f\left(x\right)$ is differentiable
in $\left(0,\infty\right)$, then for any $x\in\left(0,\infty\right)$,
we have
\begin{equation}
f'\left(x\right)<\frac{f\left(x\right)-f\left(0\right)}{x}.\label{eq:convex}
\end{equation}
\end{lem}
\begin{IEEEproof}
Fix any $x\in\left(0,\infty\right)$. According to the Lagrange mean
value theorem, there must be some $\xi\in\left(0,x\right)$ satisfying
\[
f'\left(\xi\right)=\frac{f\left(x\right)-f\left(0\right)}{x}.
\]
Since $f\left(x\right)$ is a strict concave function, then $f''\left(x\right)<0$.
Therefore, we have $f'\left(x\right)<f'\left(\xi\right)$, so that
Eq. \eqref{eq:convex} holds.
\end{IEEEproof}

With this lemma, the outage exponent can be constructed from the approximation
of the outage probability in Eq. \eqref{eq:Meijer's G approximation}.
\begin{thm}
\label{thm:Meijer exponent}For any $R$ with $\overline{R}\geq\overline{C}$,
the lower outage exponent for a parallel fading channel with $L$
subchannels is given by
\begin{equation}
p_{\mathrm{out}}\gtrsim p_{\mathrm{ex}}^{\mathrm{lower}}=e^{-L\left[\left(1-\frac{r}{L}\right)E_{1,1}\left(\gamma\right)+E_{1,0}\left(\gamma\right)+\frac{E_{0}\left(\gamma\right)}{L}\right]},\label{eq:Meijer exponent}
\end{equation}
where the functions $E_{1,1}\left(\gamma\right)$, $E_{1,0}\left(\gamma\right)$,
and $E_{0}\left(\gamma\right)$ are given by

\begin{equation}
E_{1,1}\left(\gamma\right)=\frac{\frac{\left(1+\gamma\right)^{r-1}}{\gamma^{L-1}}e^{\frac{\lambda}{\gamma}}G_{0,L}^{L,0}\left(\frac{\left(1+\gamma\right)^{r}}{\gamma^{L}}\left|\begin{array}{c}
-\\
0,0,\ldots,0
\end{array}\right.\right)}{1-e^{\frac{\lambda}{\gamma}}G_{0,L}^{L,0}\left(\frac{\left(1+\gamma\right)^{r}}{\gamma^{L}}\left|\begin{array}{c}
-\\
0,1,\ldots,1
\end{array}\right.\right)},
\end{equation}
\begin{equation}
E_{1,0}\left(\gamma\right)=\frac{1}{\gamma}E_{1,1}\left(\gamma\right),
\end{equation}
and
\begin{equation}
E_{0}\left(\gamma\right)=\frac{-\frac{\lambda}{\gamma}e^{\frac{\lambda}{\gamma}}G_{0,L}^{L,0}\left(\frac{\left(1+\gamma\right)^{r}}{\gamma^{L}}\left|\begin{array}{c}
-\\
0,1,\ldots,1
\end{array}\right.\right)}{L\left[1-e^{\frac{\lambda}{\gamma}}G_{0,L}^{L,0}\left(\frac{\left(1+\gamma\right)^{r}}{\gamma^{L}}\left|\begin{array}{c}
-\\
0,1,\ldots,1
\end{array}\right.\right)\right]}.
\end{equation}
\end{thm}
\begin{IEEEproof}
See Appendix \ref{app:Meijer exponent}.
\end{IEEEproof}

\begin{figure}[t]
\centering

\includegraphics[width=3.5in]{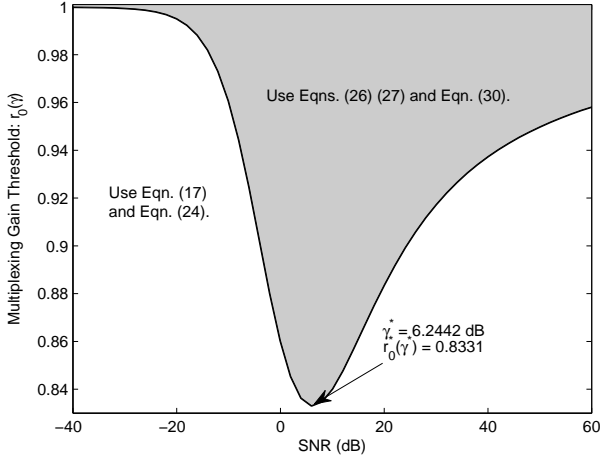}\caption{The curve of $r_{0}\left(\gamma\right)$ versus $\gamma$.\label{fig:r mean}}
\end{figure}

In Theorem \ref{thm:Meijer exponent}, the formula of the outage exponent
can only be expressed as a function of the multiplexing gain, because
$\overline{R}\geq\overline{C}$ in this case. Consider the definition
of the multiplexing gain and the condition $\overline{R}\geq\overline{C}$,
the value of $r$ must satisfy
\[
\frac{r}{L}\geq\frac{\overline{C}}{C_{\mathrm{awgn}}}=\frac{e^{\frac{1}{\gamma}}\Gamma\left(0,\gamma^{-1}\right)}{\ln\left(1+\gamma\right)}=r_{0}\left(\gamma\right).
\]
It can be shown that $r_{0}\left(\gamma\right)$ is a quasi-convex
function, and
\[
\lim_{\gamma\rightarrow0}r_{0}\left(\gamma\right)=\lim_{\gamma\rightarrow\infty}r_{0}\left(\gamma\right)=1.
\]
Fig. \ref{fig:r mean} shows the curve of $r_{0}\left(\gamma\right)$.
It can be seen that the minimum value of $r_{0}^{*}\left(\gamma\right)\approx0.8331$
is achieved at $\gamma^{*}\approx6.2442\,\mathrm{dB}$. Therefore,
for any SNR, if the multiplexing gain is below this curve, the formulas
in Theorems \ref{thm:upper bound 1} and \ref{thm:lower bound 1}
give the tight bounds; whereas if the multiplexing gain is near and
above the curve, the formulas in Theorems \ref{thm:Meijer's G bounds}
and \ref{thm:Meijer exponent} give the tight bounds.

From Theorem \ref{thm:Meijer exponent}, the finite-SNR diversity-multiplexing
tradeoff can be directly obtained as described in the following.
\begin{cor}
For a parallel fading channel with $L$ subchannels, the finite-SNR
diversity-multiplexing tradeoff is given by
\begin{equation}
\begin{aligned}d_{\mathrm{f}}^{*}\left(r,\gamma\right)= & L\left[\left(1-\frac{r}{L}\right)+\frac{1}{\gamma}\right]\frac{\left(1+\gamma\right)^{r-1}}{\gamma^{L-1}}\cdot\\
 & \frac{e^{\frac{\lambda}{\gamma}}G_{0,L}^{L,0}\left(\frac{\left(1+\gamma\right)^{r}}{\gamma^{L}}\left|\begin{array}{c}
-\\
0,0,\ldots,0
\end{array}\right.\right)}{1-e^{\frac{\lambda}{\gamma}}G_{0,L}^{L,0}\left(\frac{\left(1+\gamma\right)^{r}}{\gamma^{L}}\left|\begin{array}{c}
-\\
0,1,\ldots,1
\end{array}\right.\right)}-\\
 & \frac{\frac{\lambda}{\gamma}e^{\frac{\lambda}{\gamma}}G_{0,L}^{L,0}\left(\frac{\left(1+\gamma\right)^{r}}{\gamma^{L}}\left|\begin{array}{c}
-\\
0,1,\ldots,1
\end{array}\right.\right)}{1-e^{\frac{\lambda}{\gamma}}G_{0,L}^{L,0}\left(\frac{\left(1+\gamma\right)^{r}}{\gamma^{L}}\left|\begin{array}{c}
-\\
0,1,\ldots,1
\end{array}\right.\right)}.
\end{aligned}
\label{eq:finite SNR DMT}
\end{equation}

\end{cor}

From the finite-SNR diversity-multiplexing tradeoff, the diversity-multiplexing
tradeoff can be obtained by letting $\gamma\rightarrow\infty$. The
result is summarized in the following corollary.
\begin{cor}
\label{cor:dmt}For a parallel fading channel with $L$ subchannels,
the diversity-multiplexing tradeoff is given by
\begin{equation}
d^{*}\left(r\right)=\lim_{\gamma\rightarrow\infty}d_{\mathrm{f}}^{*}\left(r,\gamma\right)=L\left(1-\frac{r}{L}\right).\label{eq:dmt 2}
\end{equation}
\end{cor}
\begin{IEEEproof}
See Appendix \ref{app:dmt}.\end{IEEEproof}
\begin{rem}
From the proof of Theorem \ref{thm:Meijer exponent} and Corollary
\ref{cor:dmt}, we can find that if $r<L$, the outage probability
will tend to zero as SNR tends to infinity. Therefore, the diversity-multiplexing
tradeoff is valid for any $\overline{R}<C_{\mathrm{awgn}}$. As a
result, for any fixed $L$, if only $\overline{R}$ is smaller than
$C_{\mathrm{awgn}}$, the outage probability will reduce to zero as
$\gamma\rightarrow\infty$.
\end{rem}

\section{Differences between the Outage Exponent and Error Exponent\label{sec:error exponent}}

\begin{table*}[tp]
\caption{Properties of the Outage Exponent\label{tab:outage exponent}}

\centering

\begin{tabular}{|c|l|l|}
\hline 
 & \textbf{}%
\begin{tabular}{c}
\textbf{$\gamma$: }\textbf{\emph{Non-Asymptotic}}\tabularnewline
\end{tabular} & \textbf{}%
\begin{tabular}{c}
\textbf{$\gamma$: }\textbf{\emph{Asymptotic}}\tabularnewline
\end{tabular}\tabularnewline
\hline 
\textbf{}%
\begin{tabular}{c}
\textbf{$L$}\tabularnewline
\textbf{\emph{Non-Asymptotic}}\tabularnewline
\end{tabular} & %
\begin{tabular}{l}
\textbullet{} Accurate estimation of $p_{\mathrm{out}}$: Eqns. \eqref{eq:outage exponent}
\eqref{eq:upper bound 1} \eqref{eq:lower bound 1} \eqref{eq:Meijer's G bounds}.\tabularnewline
\textbullet{} Outage capacity: Eqns. \eqref{eq:outage capacity} \eqref{eq:capacity}.\tabularnewline
\textbullet{} Reliable function: Eqns. \eqref{eq:reliable func} \eqref{eq:reliable function}
\eqref{eq:Meijer exponent}.\tabularnewline
\textbullet{} Finite-SNR diversity-multiplexing tradeoff: Eqns. \eqref{eq:finite-SNR dmt def}
\eqref{eq:finite SNR DMT}.\tabularnewline
\textbullet{} SNR gain: Eqns. \eqref{eq:outage exponent} \eqref{eq:upper bound 1}
\eqref{eq:lower bound 1} \eqref{eq:Meijer's G bounds}.\tabularnewline
\end{tabular} & %
\begin{tabular}{l}
\textbullet{} diversity-multiplexing tradeoff: Eqns. \eqref{eq:dmt def}
\eqref{eq:dmt} \eqref{eq:dmt 2}.\tabularnewline
\textbullet{} If $\overline{R}<C_{\mathrm{awgn}}$, $\lim_{\gamma\rightarrow\infty}p_{\mathrm{out}}=0$:
Theorem \ref{thm:upper bound 1}.\tabularnewline
\end{tabular}\tabularnewline
\hline 
\textbf{}%
\begin{tabular}{c}
\textbf{$L$}\tabularnewline
\textbf{\emph{Asymptotic}}\tabularnewline
\end{tabular} & %
\begin{tabular}{l}
\textbullet{} Delay-limited capacity: Eqns. \eqref{eq:delay limited capacity}
\eqref{eq:dl ergodic}.\tabularnewline
\textbullet{} Ergodic capacity: Eqns. \eqref{eq:ergodic capacity}
\eqref{eq:dl ergodic}.\tabularnewline
\textbullet{} If $\overline{R}<\overline{C}$, $\lim_{L\rightarrow\infty}p_{\mathrm{out}}=0$:
Theorem \ref{thm:Meijer's G bounds}.\tabularnewline
\end{tabular} & %
\begin{tabular}{c}
---\tabularnewline
\end{tabular}\tabularnewline
\hline 
\end{tabular}
\end{table*}

In fading channels, the proposed outage exponent relates the communication
reliability, efficiency, transmission rate, SNR, and the number of
fading subchannels. In ergodic channels, the relationship among the
error probability, the transmission rate, channel noise, and the coding
lengths is characterized by a commonly used approach, referred to
as the error exponents. Recently, some works tried to generalize the
error exponent theory to fading channels. Interested readers are referred
to \cite{Shamai2002} and its references for a thorough survey on
this topic. In this section, we will discuss the differences between
our results and the theory of error exponents.

The classical results of the error exponent for parallel channels
is summarized in the following lemma \cite{Gallager1968}.
\begin{lem}
\label{lem:error exponent}Let $p_{l}\left(j|i\right),\, l=1,\ldots,L$
be the transition probability of the $l$th subchannel, and $q\left(k_{1},\ldots,k_{L}\right)$
be a probability assignment on the input vectors. If $q\left(k_{1},\ldots,k_{L}\right)=\prod_{l=1}^{L}q_{l}\left(k_{l}\right)$,
the error exponent for the parallel fading channel is then given by
the exponentially tight upper bound of the error probability:
\[
p_{\mathrm{err}}\leq e^{-N\sum_{l=1}^{L}E_{r}^{l}\left(\rho\right)},
\]
where
\[
E_{r}^{l}\left(\rho\right)=\max_{\bm{q}_{l}}\left\{ E_{o}^{l}\left(\rho,\bm{q}_{l}\right)-\rho\frac{\partial}{\partial\rho}E_{o}^{l}\left(\rho,\bm{q}_{l}\right)\right\} ,
\]
and
\[
R\left(\rho\right)=\sum_{l=1}^{L}R_{l}\left(\rho\right)=\sum_{l=1}^{L}\frac{\partial}{\partial\rho}E_{o}^{l}\left(\rho,\bm{q}_{l}\right).
\]
The function $E_{o}^{l}\left(\rho,\bm{q}_{l}\right)$ is given by
\[
E_{o}^{l}\left(\rho,\bm{q}_{l}\right)=-\ln\sum_{j=0}^{J-1}\left(\sum_{k=0}^{K-1}q_{l}\left(k\right)p_{l}\left(j|k\right)^{\frac{1}{1+\rho}}\right)^{1+\rho}.
\]
The parameter $\rho$ is the magnitude of the slope of the $E_{r}^{l}\left(\rho\right)$
versus $R\left(\rho\right)$ curve.
\end{lem}

From Lemma \ref{lem:error exponent}, if $p_{l}\left(i|j\right),\, l=1,\ldots,L$
are the same for any subchannel, the error exponent will be reduced
to
\begin{equation}
p_{\mathrm{err}}\leq e^{-L\left(NE_{r}^{l}\left(\rho\right)\right)},\label{eq:error exponent}
\end{equation}
and 
\begin{equation}
R\left(\rho\right)=LR_{l}\left(\rho\right)=L\frac{\partial}{\partial\rho}E_{o}^{l}\left(\rho,\bm{q}_{l}\right).\label{eq:error exponent rate}
\end{equation}
Therefore, if and only if $R_{l}<C_{\mathrm{awgn}}$, $p_{\mathrm{err}}$
will tend to zero when $L\rightarrow\infty$ for a fixed $N$. It
can be shown that $E_{r}^{l}\left(\rho\right)$ in Eq. \eqref{eq:error exponent}
is just the error exponent for the $l$th subchannel.

By comparing Eqns. \eqref{eq:error exponent} \eqref{eq:error exponent rate}
and Eq. \eqref{eq:upper bound 1}, it can be found that the outage
exponent and error exponent have a similar form. However, they are
different. First of all, the outage exponent considers the non-ergodic
parallel fading channel, i.e., each subchannel is associated with
a random channel gain; whereas each subchannel in the error exponent
is ergodic. The results that tried to generalize the error exponent
method to fading channels, however, can only be applied to ergodic
fading channels, e.g., the fully interleaved block-fading channel
in \cite{Sason2001,Lun2002}. For the non-ergodic parallel fading
channel, the transition probability of each subchannel requires knowledge
of the channel, i.e., we need $p_{l}\left(j|i,h_{l}\right)$. Then,
from the perspective of error exponent, the outage probability is
given by 
\begin{equation}
p_{\mathrm{out}}=\Pr\left\{ \sup\left\{ R\left(\rho\right):\sum_{l=1}^{L}E_{r}^{l}\left(\rho,\bm{h}\right)>0\right\} <R\right\} .\label{eq:error exponent outage}
\end{equation}
Because $E_{r}^{l}\left(\rho,\bm{h}\right)$ must be maximized over
$\bm{q}_{l}$ for every sample of $\bm{h}$, it is very difficult
to get a closed-form formula for the outage probability. Some attempts
can be found in \cite{Divsalar2000,Wu2007}. By comparing Eq. \eqref{eq:outage definition}
and Eq. \eqref{eq:error exponent outage}, it can be found that:
1) the outage exponent only focuses on the effect of channel fading;
and 2) all of the details about the channel coding are dropped by
assuming that an ideal coding scheme has been adopted. This is reasonable
because the AWGN channel capacity can be approximately achieved by
LDPC and Turbo codes. Actually, this is just the basic idea in \cite{Ozarow1994}.

Another important difference is that the error exponent for the parallel
fading channel is $L$ times the error exponents of each subchannel,
as shown in Eq. \eqref{eq:error exponent}. For the outage exponent,
however, the reliable function defined in Eq. \eqref{eq:reliable func}
is not the outage exponent for one subchannel, since the outage probability
for each subchannel is given by
\[
p_{\mathrm{out}}^{l}=1-e^{-\frac{e^{\overline{R}}-1}{\gamma}}.
\]
Therefore, we can independently study the subchannels of an ergodic
parallel fading channel. For the non-ergodic parallel fading channel,
we have to view the channel as a whole so as to study its performance.
To illustrate the differences more clearly, the obtained results in
this paper for the outage exponent are briefly summarized in Table
\ref{tab:outage exponent}.

\section{Discussion on Coding Issues for the Parallel Fading Channel\label{sec:coding}}

According to the above results, if the transmitter has perfect knowledge
of CSI of each subchannel, the outage exponent can be achieved easily.
As a matter of fact, the outage performance can be better, in the
sense of coding gain, when the water-filling power allocation is applied.
If the transmitter has no CSI, however, how to achieve the outage
performance as good as possible is a key problem in this context.
By recalling the definition of the outage probability given in Eq.
\eqref{eq:outage definition}, the optimal outage performance can
be achieved if the coding scheme with rate $R$ is capable of achieving
the capacity of each subchannel without CSI at the transmitter. This
section will discuss some basic issues in designing the best codes
for the parallel fading channel.

It is known that the minimum distance between two codewords determines
the decode error probability in the AWGN channel \cite{Proakis2007}.
In fading channels, another key parameter is the minimum product distance
of two codewords \cite{Tse2005}. In brief, the minimum distance determines
the capability of codes to combat noise; whereas the minimum product
distance determines the capability to combat channel fading. Eq.
\eqref{eq:outage definition} indicates that the outage probability
is determined by channel fading. Therefore, the minimum product distance
is the key metric to evaluate a coding scheme for the parallel fading
channel.

In signal space, each codeword can be seen as a vector or point in
the $L$-dimension space, i.e., $L$-dimension constellation. Moreover,
the channel gains on $L$ subchannels are orthogonal random variables,
because they are independent with zero means. Therefore, the $l$th
axis of the $L$-dimension constellation corresponds to the channel
gain of the $l$th subchannel (i.e., $h_{l}$). Let $\bm{X}_{A}=\left(\bm{x}_{A1},\bm{x}_{A2},\ldots,\bm{x}_{AL}\right)^{\mathrm{T}}$
and $\bm{X}_{B}=\left(\bm{x}_{B1},\bm{x}_{B2},\ldots,\bm{x}_{BL}\right)^{\mathrm{T}}$
denote the codewords for the information $A$ and $B$, respectively.
The normalized product distance of $\bm{X}_{A}$ and $\bm{X}_{B}$
is then defined by
\begin{equation}
D_{p}=\frac{1}{\sqrt{\gamma}}\prod_{l=1}^{L}\left|\bm{x}_{Al}-\bm{x}_{Bl}\right|.\label{eq:product distance}
\end{equation}
Therefore, the coding scheme must be designed to maximize the minimum
value of $D_{p}$, denoted by $D_{p}^{\mathrm{min}}$, so as to optimize
the outage performance. If $\bm{x}_{Al}=\bm{x}_{Bl}$ for some $l$,
there must be a hyperplane which is orthogonal with the $l$th axis
such that $\bm{x}_{Al}-\bm{x}_{Bl}$ is parallel with this hyperplane.
Clearly, if all the channel gains on that hyperplane are very small,
the receiver cannot distinguish $\bm{X}_{A}$ from $\bm{X}_{B}$.
In other words, we lose one dimension, i.e., the $l$th subchannel,
to combat the channel fading. As a matter of fact, the number of terms
that $\bm{x}_{Al}\neq\bm{x}_{Bl},\, l=1,2,\ldots,L$ in Eq. \eqref{eq:product distance}
is the diversity order achieved by this signal constellation \cite{Oggier2004}.

To the best of our knowledge, there are two main approaches to maximize
the minimum product distance. The first class is the rotated $\mathbb{Z}^{L}$-lattices
code, which is based on the algebraic number theory and lattices theory.
The basic idea is to construct the $L$-dimension constellation with
the size $2^{R}$ based on $\mathbb{Z}^{L}$-lattices. Then, the constellation
will be rotated to an appropriate angle such that $D_{p}^{\min}$
is maximized. According to the lattices theory, $D_{p}^{\min}$ can
be calculated theoretically. One can refer to \cite{Wang2003,Oggier2004}
and its references for detailed discussions on this coding scheme.
Another advantage of the rotated $\mathbb{Z}^{L}$-lattices code lies
in the fact that it can reduce the peak-to-average power ratio (PAPR)
\cite{Henkel2000}. As a matter of fact, the single carrier frequency
division multiple access (SC-FDMA) can be seen as a special case of
rotated $\mathbb{Z}^{L}$-lattices codes, where the rotated angle
is determined by the discrete Fourier transform (DFT) matrix. This
rotation can increase $D_{p}^{\mathrm{min}}$ or the distance between
two points with zero product distance. Therefore, compared to uncoded
orthogonal frequency division multiple access (OFDMA) systems, SC-FDMA
has a better performance in form of decoding error probability and
PAPR \cite{Bai2010a}. Another approach is the permutation code which
was proposed to achieve the optimal diversity-multiplexing tradeoff
\cite{Tse2005,Tavildar2006}. The basic idea is to use the constellation
of size $2^{R}$ on a complex plane for each subchannel. The points
in the constellation for each subchannel are permuted, so that the
product distance is maximized. The permutation operation on each subchannel
can be seen as the product of the original information times a specific
matrix, which is referred to as the universal decodable matrix (UDM).
Based on the Pascal triangle, the universal decodable matrices can
be constructed directly for every subchannel \cite{Ganesan2007}.
From the perspective of rotated $\mathbb{Z}^{L}$-lattices code, the
permutation code can be seen as a way to construct a $L$-dimension
constellation with size $2^{LR}$. The permutation rules imply that
we should choose $2^{R}$ points from $2^{LR}$, so that the product
distance is maximized. It is not trivial to evaluate the decoding
error probability of the permutation code, because the analytic formula
for $D_{p}^{\min}$ has not been obtained. According to the construction
process of the permutation code, the PAPR performance could be worse
than the rotated $\mathbb{Z}^{L}$-lattices code for the same average
power. Another important difference is that the constellation size
of each subchannel is $2^{\overline{R}}=2^{R/L}$ for the rotated
$\mathbb{Z}^{L}$-lattices code, while it is $2^{R}$ for the permutation
code.

It is possible to combine the advantages of the rotated $\mathbb{Z}^{L}$-lattices
code and permutation code. First, we can construct a $L$-dimension
constellation with the size of $M$ ($2^{R}\leq M$). Then, the constellation
can be rotated to an appropriate angle to maximize $D_{p}^{\min}$.
After that, $2^{R}$ points from $M$ can be chosen to further maximize
$D_{p}^{\min}$. In order to find the optimal coding scheme, we consider
the asymptotic case that $M\rightarrow\infty$. Therefore, the $L$-dimension
constellation becomes a continuos $L$-dimension hypersphere whose
radius is determined by the peak power of the coding scheme. Then,
the optimal coding is to find the coordinates for $2^{R}$ points
which maximize $D_{p}^{\min}$ under the constraint of average power.
This is an interesting idea that formulates the optimal coding design
problem, which is often seen as a complicated discrete optimization
problem, into an optimization problem on continuos variables. However,
the general approach, which is out of the scope of this paper, still
needs to be investigated in depth.

\section{Numerical Results\label{sec:numerical}}

In this section, some numerical results will be presented to verify
the theoretical derivation in the previous sections. According to
the proof of the proposed theorems, the accuracy of the bounds will
be tight as the number of subchannels increases. Hence, we only need
to compare the results for small $L$. Since $\overline{C}$ is a
function of SNR, the average multiplexing gain (or normalized target
rate) $\overline{r}=\frac{r}{L}=\frac{\overline{R}}{C_{\mathrm{awgn}}}$
is used to guarantee $\overline{R}<\overline{C}$ or $\overline{R}\geq\overline{C}$.

\begin{figure}[t]
\centering

\includegraphics[width=3.5in]{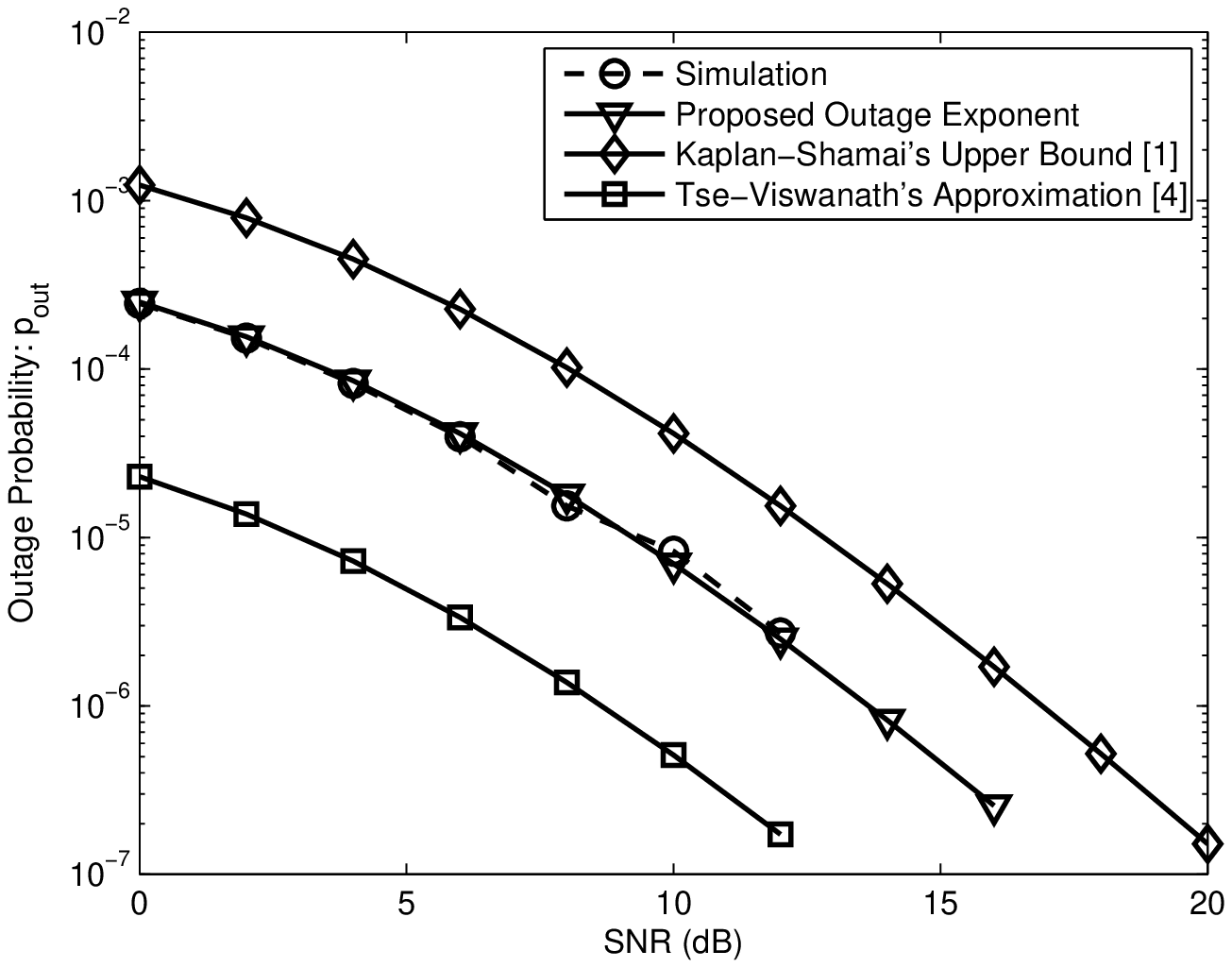}

\caption{Outage probability comparison for $L=4$ and $\overline{r}=0.1$.\label{fig:accurate 1}}
\end{figure}

The first set of simulations is to verify the accuracy of the proposed
upper and lower bounds. In the simulation, we first generate the samples
of the parallel fading channel, and then compare the instantaneous
channel capacity with the target transmission rate, which yields the
simulation results for the outage probability. Fig. \ref{fig:accurate 1}
compares the simulation results and the theoretical results of outage
exponent for the $\overline{R}<\overline{C}$ case. In Fig. \ref{fig:accurate 1},
the number of subchannels is $L=4$, and the average multiplexing
gain is $\overline{r}=0.1$. The outage exponent is computed through
Eq. \eqref{eq:upper bound 1}. It can be seen that the proposed outage
exponent is nearly identical with the simulation results in the full
SNR range. The Kaplan-Shamai's upper bound in Eq. \eqref{eq:outage bound Shamai}
and the Tse-Viswanath's approximation in Eq. \eqref{eq:outage bound Tse}
have the accurate slope but not the intercept in this case.

\begin{figure}[t]
\centering

\includegraphics[width=3.5in]{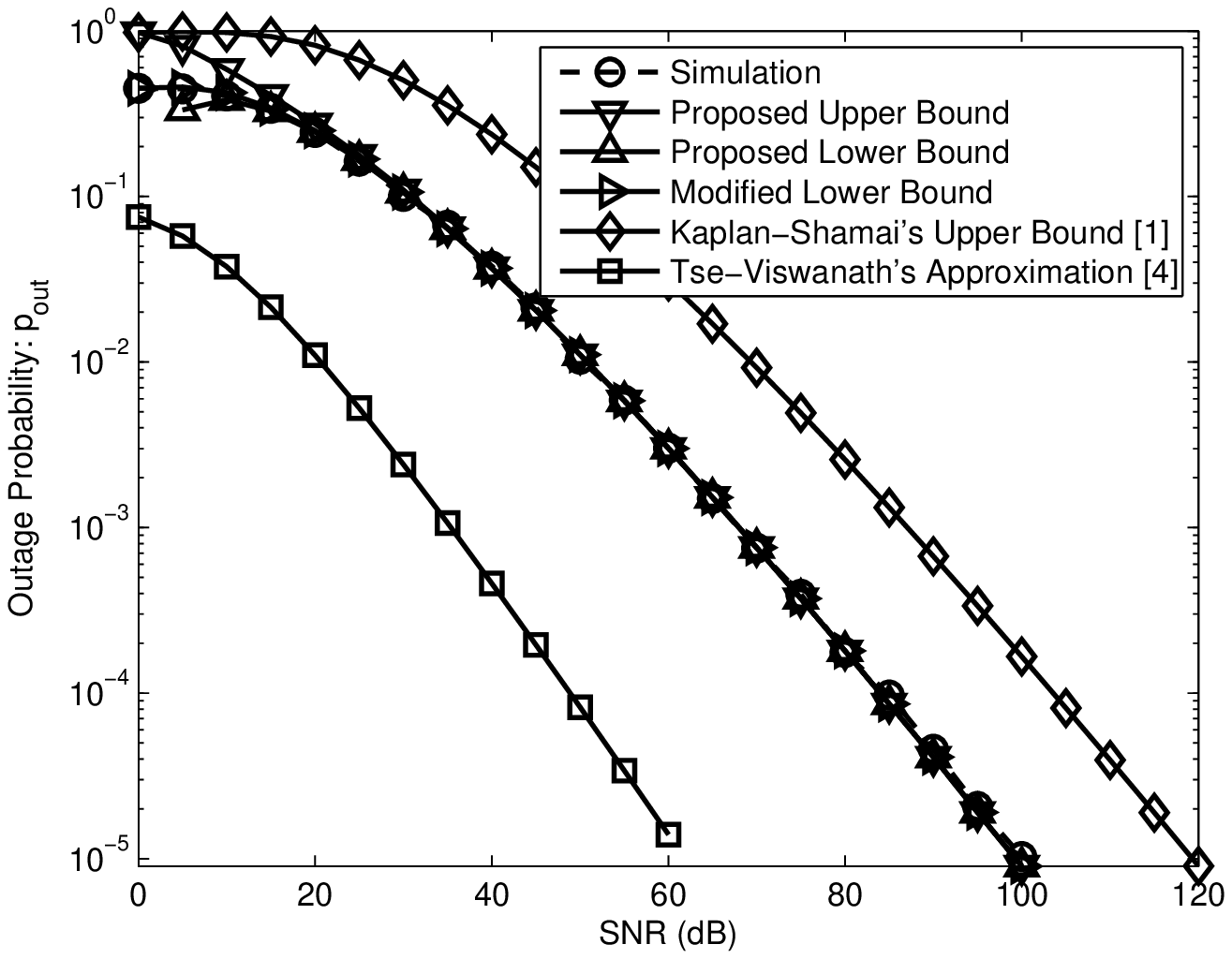}

\caption{Outage probability comparison for $L=4$ and $\overline{r}=0.8$.\label{fig:accurate 2}}
\end{figure}

For Fig. \ref{fig:accurate 2}, we compare the simulation results
and the outage exponent for the $\overline{R}\geq\overline{C}$ case.
The proposed upper and lower bounds are calculated through Eq. \eqref{eq:Meijer's G bounds}.
It can be seen that the proposed upper and lower bounds will converge
to the simulation values as SNR increases. Because the proposed lower
bound is only valid in the high SNR regime, the point at $\unit[0]{dB}$
of the curve is lacking. Hence, the modified lower bound in Proposition
\ref{pro:Meijer's G approximation} has been proposed to overcome
this shortcoming. In Fig. \ref{fig:accurate 2}, it can be seen that
the modified lower bound is an accurate estimation for the outage
probability. Clearly, the upper bound in Eq. \eqref{eq:outage bound Shamai}
cannot provide an accurate estimation of the outage probability. The
approximation in Eq. \eqref{eq:outage bound Tse} does not have a
good accuracy either, since it can only estimate the slope of the
outage probability in high SNRs for large average multiplexing gains.

\begin{figure}[t]
\centering

\includegraphics[width=3.5in]{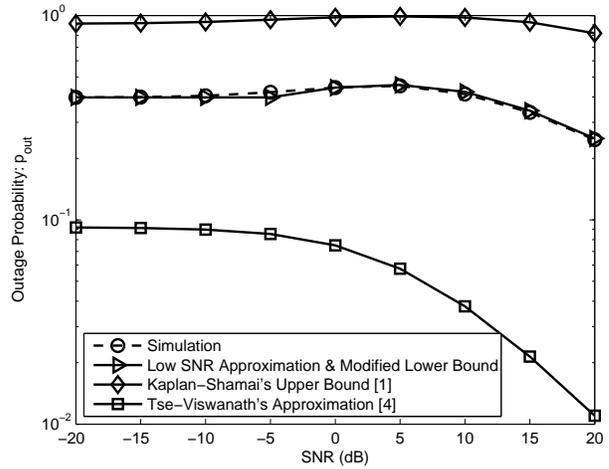}

\caption{Outage probability comparison for $L=4$ and $\overline{r}=0.8$ in
the low SNR regime.\label{fig:lowSNR}}
\end{figure}

Fig. \ref{fig:lowSNR} shows the simulation results of the outage
probability and the proposed bounds from $\unit[-20]{dB}$ to $\unit[20]{dB}$.
It can be seen that the low SNR approximation in Eq. \eqref{eq:Meijer's G_lower},
which is ploted from $\unit[-20]{dB}$ to $\unit[-5]{dB}$, is very
tight as expected. From $0$ to $\unit[20]{dB}$, the proposed approximation
in Proposition \ref{pro:Meijer's G approximation} has been used,
which is also very tight. The Kaplan-Shamai's upper bound and Tse-Viswanath's
approximation are also plotted for comparison. In the Kaplan-Shamai's
upper bound, i.e., Eq. \eqref{eq:outage bound Shamai}, the large
deviation principle is applied. Therefore, the slope is accurate when
$\overline{R}<\overline{C}$ or $\overline{R}>\overline{C}$, and
is not accurate enough when $\overline{R}$ is close to $\overline{C}$.
According to Fig. \ref{fig:r mean}, when SNR is between $\unit[0]{dB}$
and $\unit[10]{dB}$, $\overline{r}=0.8$ is very close to the threshold
curve $r_{0}$. In this range, Kaplan-Shamai's upper bound converges
to $1$. When SNR is smaller than $\unit[0]{dB}$ or larger than $\unit[10]{dB}$,
the difference between $\overline{r}=0.8$ and the threshold curve
in Fig. \ref{fig:r mean} is considerably large. Therefore, the Kaplan-Shamai's
upper bound characterizes the accurate slope in this SNR range.

\begin{figure}[t]
\centering

\includegraphics[width=3.5in]{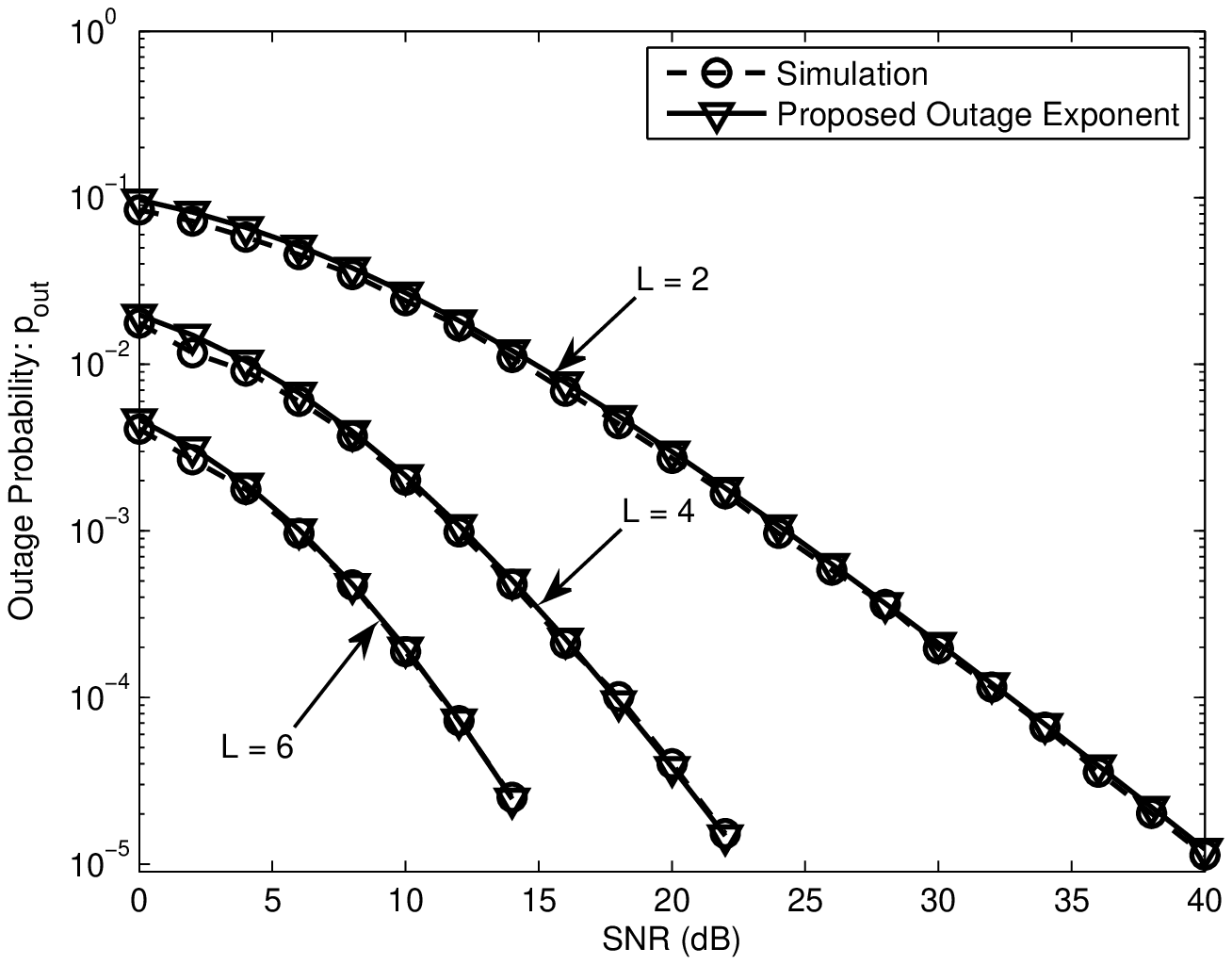}

\caption{Outage probability comparison for $L=2,4,6$, and $\overline{r}=0.3$.\label{fig:outage 1}}
\end{figure}

\begin{figure}[t]
\centering

\includegraphics[width=3.5in]{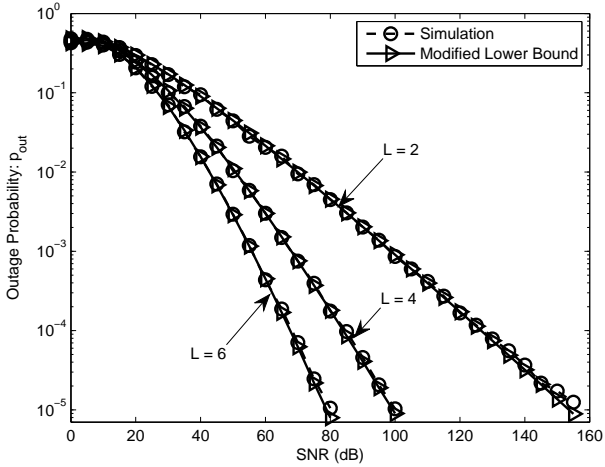}

\caption{Outage probability comparison for $L=2,4,6$, and $\overline{r}=0.8$.\label{fig:outage 2}}
\end{figure}

The second set of simulations illustrate the behavior of the outage
probability as the number of subchannels increases. Figs. \ref{fig:outage 1}
and \ref{fig:outage 2} show the simulated outage probability and
the theoretical results for $\overline{R}<\overline{C}$ and $\overline{R}\geq\overline{C}$,
respectively. The number of subchannels $L$ is set to be $2$, $4$,
and $6$, respectively. As expected, the proposed bounds are accurate
even for $L=2$. From Fig. \ref{fig:outage 1}, it can be found that
the slopes of the outage probability for $L=2,4,6$ do not vary much
with one another. The performance gain are mainly comes from the SNR
gain. This phenomenon can be explained from Eq. \eqref{eq:upper bound 1}.
Clearly, for a fixed $L$, we have
\[
\begin{aligned}p_{\mathrm{out}} & \lesssim\frac{1}{\sqrt{2\pi L}\sigma\Xi\left(0\right)}e^{-LE_{1}\left(R,\gamma\right)}\end{aligned}
.
\]
Therefore, varying $L$ results in a great change in the coefficient
of the exponential function. However, as shown in Fig. \ref{fig:outage 2},
the performance gains are mainly determined by the diversity gain
for the $\overline{R}\geq\overline{C}$ case. This phenomenon occurs
because the coefficient in Eq. \eqref{eq:Meijer exponent} is one
for any $L$. This results indicate that the power allocation is very
important for the small average multiplexing gain case, but may not
result in a significant performance gain in the large average multiplexing
gain case.

\begin{figure}[t]
\centering

\includegraphics[width=3.5in]{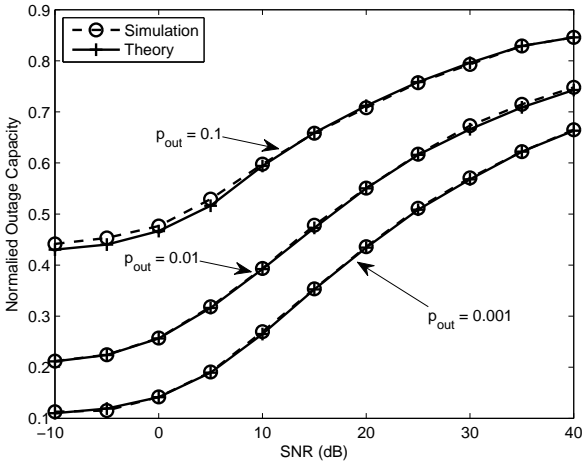}

\caption{Outage capacity comparison for $p_{\mathrm{out}}=0.001,0.01,0.1$,
and $L=4$.\label{fig:capacity}}
\end{figure}

The outage capacity defined as the maximum transmission rate which
guarantee the outage probability is smaller than a target value. Fig.
\ref{fig:capacity} shows the comparison between the simulated outage
capacity and theoretical results. The normalized outage capacity in
the figure is just the maximum average multiplexing gain which can
be achieved for a given outage probability. In the simulation, the
binary search based ordinal optimization method is used to calculate
$C_{\varepsilon}$ \cite{Jia2006}. First, fix a desired outage probability
$p_{\mathrm{out}}$, and let $C_{\mathrm{norm}}=C_{\mathrm{norm}}^{\mathrm{max}}=2$
and $C_{\mathrm{norm}}^{\mathrm{min}}=0$. If we set $R=C_{\mathrm{norm}}\cdot C_{\mathrm{awgn}}$
and give a SNR value, the simulated outage probability $\hat{p}_{\mathrm{out}}$
can then be obtained. If $|\hat{p}_{\mathrm{out}}-p_{\mathrm{out}}|<\delta$,
$C_{\mathrm{norm}}$ is seen as the correct value. However, if $|\hat{p}_{\mathrm{out}}-p_{\mathrm{out}}|\geq\delta$,
$C_{\mathrm{norm}}^{\mathrm{max}}$ and $C_{\mathrm{norm}}^{\mathrm{min}}$
will be changed with the following rules:
\begin{enumerate}
\item if $\hat{p}_{\mathrm{out}}-p_{\mathrm{out}}\geq\delta$, set $C_{\mathrm{norm}}^{\mathrm{max}}=C_{\mathrm{norm}}$
and go to 3);
\item if $\hat{p}_{\mathrm{out}}-p_{\mathrm{out}}\leq-\delta$, set $C_{\mathrm{norm}}^{\mathrm{min}}=C_{\mathrm{norm}}$;
\item set $C_{\mathrm{norm}}=\frac{1}{2}\left(C_{\mathrm{norm}}^{\mathrm{max}}+C_{\mathrm{norm}}^{\mathrm{min}}\right)$.
\end{enumerate}
After the correct $C_{\mathrm{norm}}$ is obtained, the SNR will be
changed to a new value. The normalized $\varepsilon$-outage capacity
at any SNR value can then be obtained. In our simulations, $\delta$
is set to $\frac{1}{1000}p_{\mathrm{out}}$. The corresponding theoretical
results are computed through Eq. \eqref{eq:capacity}, since $r_{0}\left(\gamma\right)\rightarrow1$
as $\gamma\rightarrow\infty$. From Fig. \ref{fig:capacity}, it can
be seen that the theoretical results are very accurate for $p_{\mathrm{out}}=0.001$
and $p_{\mathrm{out}}=0.01$. For $p_{\mathrm{out}}=0.1$, the two
curves have a small gap in the low SNR regime, because $r_{0}\left(\gamma\right)$
achieves the minimum value at $\gamma^{*}\approx6.2442\,\mathrm{dB}$.

\begin{figure}[t]
\centering

\includegraphics[width=3.5in]{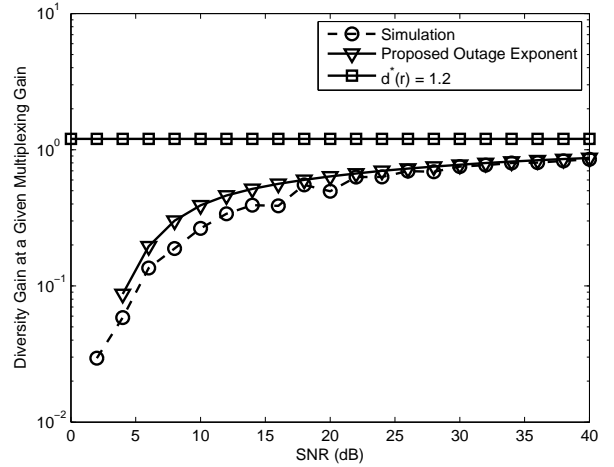}

\caption{Diversity gain at a given average multiplexing gain for $L=4$ and
$\overline{r}=0.7$.\label{fig:diversity 1}}
\end{figure}

\begin{figure}[t]
\centering

\includegraphics[width=3.5in]{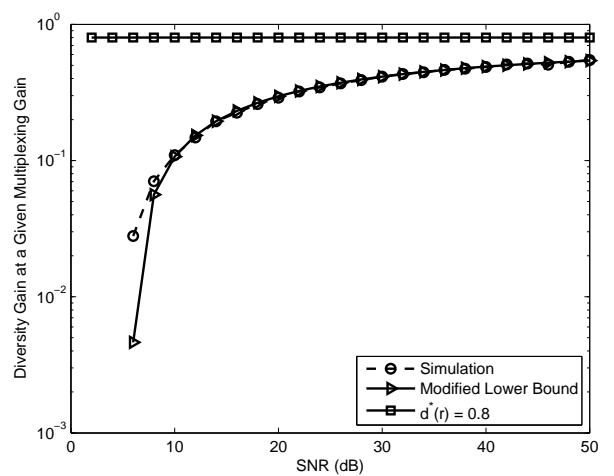}

\caption{Diversity gain at a given average multiplexing gain for $L=4$ and
$\overline{r}=0.8$.\label{fig:diversity 2}}
\end{figure}

The last set of simulations is to illustrate the diversity gain versus
SNR for a given average multiplexing gain. For $\overline{R}<\overline{C}$,
as shown in Fig. \ref{fig:diversity 1}, the average multiplexing
gain is set to $\overline{r}=0.7$. The simulated diversity gain is
obtained by differentiating the outage probability. The theoretical
results is computed by Eq. \eqref{eq:upper bound 1}. The asymptotic
diversity gain is also plotted for comparison. It can be seen that
the theoretical and simulation curves have a small gap when SNR is
smaller than $20\,\mathrm{dB}$. This is because $\overline{r}=0.7$
is close to the minimum value of $r_{0}\left(\gamma\right)$ in that
SNR range, which yields a loose large deviation bound. Fig. \ref{fig:diversity 2}
shows the diversity at a given average multiplexing gain for $\overline{r}=0.8$.
In contrast to Fig. \ref{fig:diversity 1}, the theoretical curve
in this figure is calculated by Eq. \eqref{eq:finite SNR DMT}. Although
$\overline{r}=0.8$ is still smaller than the minimum value of $r_{0}\left(\gamma\right)$,
the proposed bound is nearly identical with the simulation results.
Therefore, the proposed bound for $\overline{R}\geq\overline{C}$
is also very tight for $\overline{r}$ is of the middle values. From
these figures, the finite-SNR diversity gain approaches to the asymptotic
value as SNR increases. Therefore, for high SNRs, such as more than
$40\,\mathrm{dB}$, the asymptotic diversity-multiplexing tradeoff
can be used to evaluate the performance instead of the corresponding
finite-SNR diversity-multiplexing tradeoff.

\begin{rem}
Throughout this section, the numerical evaluation of the theoretical
results need to compute the Meijer's $G$-function. However, the numerical
computation problem of Meijer's $G$-function has not been fully solved.
MATLAB R2010a does not have this function at all, while the realizations
in Mathematica 7.0 and Maple 14 have severe bugs in some cases. From
a large number of experiments, for instance, Mathematica 7.0 will
give wrong results if the input parameters are decimal and not fractional.
The designers for Mathematica, who wrote the program of Meijer\textquoteright{}s
G-function, also pointed out the design problems of this function,
especially the logarithmic cases \cite{Adamchik1995}. Therefore,
a software package of Meijer's G-function is developed for MATLAB
in \cite{Shi2010}. In the simple poles case, the algorithm can go
straightforwardly to the generalized hyper-geometric functions. Meanwhile,
the algorithm can also deal with some cases which Maple 14 cannot
handle. For the super complicated and potentially buggy cases, i.e.,
the logarithmic cases, it is very difficulty to design a general algorithm.
Aiming at this problem, the accurate formulas to evaluate the residues
of the integration have been derived, and can be used to numerically
evaluate Meijer's $G$-function with arbitrary precision.
\end{rem}

\section{Conclusions\label{sec:conclusions}}

This paper focused on the parallel fading channel and proposed a unified
performance metric, referred to as the outage exponent. The outage
exponent is defined as the exponentially tight upper and lower bounds
of the outage probability for any number of subchannels, any SNR,
and any target transmission rate. Based on the latest results in large
deviations theory, Meijer's $G$-function, and the method of integral
around a contour, the outage exponent is obtained for both $\overline{R}<\overline{C}$
and $\overline{R}\geq\overline{C}$ cases. From the accurate estimation
of the outage probability, the reliable function, outage capacity,
finite-SNR diversity-multiplexing tradeoff, SNR gain, and also the
asymptotic performance metrics, including the delay-limited capacity,
ergodic capacity, and diversity-multiplexing tradeoff have been calculated.
In order to achieve the proposed outage exponent, the coding schemes
which maximize the minimum product distance have also been discussed.
Therefore, it can be concluded that the proposed outage exponent framework
provides a powerful tool for analyzing and evaluating the performance
of existing and upcoming communication systems.

\section*{Acknowledgment}

The authors would like to thank Professor John S. Sadowsky for insightful
discussion on large deviations theory.

\appendices{}

\section{Proof of Theorem \ref{thm:upper bound 1}\label{app:upper bound 1}}

Define a sequence of random variables $\left\{X_{l}:l=1,\ldots,L\right\}$
by letting
\[
X_{l}=\overline{R}-\ln\left(1+\left|h_{l}\right|^{2}\gamma\right).
\]
We further let $Y_{L}=\sum_{l=1}^{L}X_{l}$, then the outage probability
defined in Eq. \eqref{eq:outage definition} is equivalent to
\[
p_{\mathrm{out}}=\Pr\left\{ \frac{1}{L}\sum_{l=1}^{L}X_{l}>0\right\} =\Pr\left\{ \frac{1}{L}Y_{L}>0\right\} .
\]

Because the elements of $\left\{ h_{l}:l=1,\ldots,L\right\} $ are
independent with the identical distribution of $\mathcal{CN}\left(0,1\right)$,
the logarithmic moment generating function of $Y_{L}$ is given by
\[
\begin{aligned}\Lambda_{L}\left(\xi\right) & =\ln\mathsf{E}\left\{ e^{\xi Y_{L}}\right\} =\ln\left(e^{\xi\overline{R}}\mathsf{E}\left\{ e^{-\xi\ln\left(1+\left|h_{l}\right|^{2}\gamma\right)}\right\} \right)^{L}\\
 & =L\left[\xi\overline{R}+\ln\left(\int_{0}^{\infty}e^{-\xi x}\frac{1}{\mathsf{\gamma}}e^{-\frac{e^{x}-1}{\mathsf{\gamma}}+x}dx\right)\right]\\
 & =L\left[\xi\overline{R}+\ln\left(e^{\frac{1}{\mathsf{\gamma}}}\frac{1}{\mathsf{\gamma}}\int_{1}^{\infty}t^{-\xi}e^{-\frac{t}{\gamma}}dt\right)\right]\\
 & =L\left[\left(\overline{R}-\ln\gamma\right)\xi+\ln\left(e^{\frac{1}{\gamma}}\Gamma\left(1-\xi,\gamma^{-1}\right)\right)\right].
\end{aligned}
\]
Therefore, we have
\[
\begin{aligned}\Lambda\left(\xi\right) & =\lim_{L\rightarrow\infty}\frac{1}{L}\Lambda_{L}\left(\xi\right)\\
 & =\left(\overline{R}-\ln\gamma\right)\xi+\ln\left(e^{\frac{1}{\gamma}}\Gamma\left(1-\xi,\gamma^{-1}\right)\right).
\end{aligned}
\]
Notice that $\overline{R}<\overline{C}$, i.e., $\mathsf{E}\left\{ Y_{L}\right\} <0$,
the Legendre-Fenchel transform of $\Lambda\left(\xi\right)$ at $s=0$
is then given by
\begin{equation}
\begin{aligned} & \Lambda^{*}\left(0\right)=\sup_{\xi\in\mathbb{R}}\left\{ -\Lambda\left(\xi\right)\right\} \\
= & -\inf_{\xi\geq0}\left\{ \left(\overline{R}-\ln\gamma\right)\xi+\ln\left(e^{\frac{1}{\gamma}}\Gamma\left(1-\xi,\gamma^{-1}\right)\right)\right\} \\
= & \left(\ln\gamma-\overline{R}\right)\Xi\left(0\right)-\ln\left(e^{\frac{1}{\gamma}}\Gamma\left(1-\Xi\left(0\right),\gamma^{-1}\right)\right)\\
= & \left(C_{\mathrm{awgn}}-\overline{R}\right)\Xi\left(0\right)-\Xi\left(0\right)\ln\left(1+\frac{1}{\gamma}\right)\\
 & -\ln\left(e^{\frac{1}{\gamma}}\Gamma\left(1-\Xi\left(0\right),\gamma^{-1}\right)\right)
\end{aligned}
\label{eq:Legendre-Fenchel}
\end{equation}
Since $\Lambda\left(\xi\right)$ is convex and differentiable, $\xi=\Xi\left(0\right)$
is the solution of the following equation:
\[
\frac{\partial}{\partial\xi}\Lambda\left(\xi\right)=0.
\]
By computing the derivative, we have
\[
\frac{\partial}{\partial\xi}\Lambda\left(\xi\right)=\overline{R}-\frac{1}{\Gamma\left(1-\xi,\gamma^{-1}\right)}G_{2,3}^{3,0}\left(\frac{1}{\gamma}\left|\begin{aligned} & 1,1\\
 & 0,0,1-\xi
\end{aligned}
\right.\right).
\]
Therefore, $\Xi\left(0\right)$ is the solution of Eq. \eqref{eq:Xi0}.
According to the results in \cite{Bucklew1993}, $\sigma^{2}$ is
given by
\[
\begin{aligned}\sigma^{2}= & \left.\frac{\partial^{2}}{\partial\xi^{2}}\Lambda\left(\xi\right)\right|_{\xi=\Xi\left(0\right)}\\
= & \frac{2}{\Gamma\left(1-\Xi\left(0\right),\gamma^{-1}\right)}G_{3,4}^{4,0}\left(\frac{1}{\gamma}\left|\begin{aligned} & 1,1,1\\
 & 0,0,0,1-\Xi\left(0\right)
\end{aligned}
\right.\right)-\\
 & \left[\frac{1}{\Gamma\left(1-\Xi\left(0\right),\gamma^{-1}\right)}G_{2,3}^{3,0}\left(\frac{1}{\gamma}\left|\begin{aligned} & 1,1\\
 & 0,0,1-\Xi\left(0\right)
\end{aligned}
\right.\right)\right]^{2}.
\end{aligned}
\]
Clearly, we have

\[
\Lambda_{L}^{*}\left(0\right)=\sup_{\xi\geq0}\left\{ -\Lambda_{L}\left(\xi\right)\right\} =L\Lambda^{*}\left(0\right).
\]

According to the proof of Cram\'er's theorem in Chapter 2.2.1 of
\cite{Dembo1998}, the following inequality holds for any $L\in\mathbb{N}$:
\[
p_{\mathrm{out}}\leq\psi e^{-L\Lambda^{*}\left(0\right)}=p_{\mathrm{ex}}^{\mathrm{upper}}.
\]
where $\psi$ is a slowly varying function. As a matter of fact, from
Remark (c) on Cram\'er's theorem in Chapter 2.2.1 of \cite{Dembo1998},
a loose estimation of $\psi$ is $2$. Let $\mathcal{F}=\left\{ Y_{L}:\frac{1}{L}Y_{L}\geq0\right\} $
and $\mathcal{G}=\left\{ Y_{L}:\frac{1}{L}Y_{L}>0\right\} $ with
$\mathcal{G}\subseteq\mathcal{F}$, then from Cram\'er's theorem,
we have 
\[
\begin{aligned}-\Lambda^{*}\left(0\right) & \leq\liminf_{L\rightarrow\infty}\frac{1}{L}\ln\Pr\left\{ \mathcal{G}\right\} \\
 & \leq\limsup_{L\rightarrow\infty}\frac{1}{L}\ln\Pr\left\{ \mathcal{F}\right\} \leq-\Lambda^{*}\left(0\right),
\end{aligned}
\]
which implies that 
\[
\lim_{L\rightarrow\infty}\frac{1}{L}\ln p_{\mathrm{out}}=-\Lambda^{*}\left(0\right).
\]
Therefore, $\Lambda^{*}\left(0\right)$ is a good rate function in
the sense of large deviation, and there must be a slowly varying function
$\varphi$ with $\varphi<\psi$ satisfying
\[
p_{\mathrm{ex}}^{\mathrm{lower}}=\varphi e^{-L\Lambda^{*}\left(0\right)}\leq p_{\mathrm{out}}.
\]
The above arguments can also be verified by the derivations and results
in two important literatures on probability inequalities and large
deviation results for sums of independent random variables \cite{Hoeffding1963,Nagaev1979}.

In order to obtain a tight upper bound, the only needed work is to
estimate $\psi$ accurately. Because the elements of $\left\{ X_{l}:l=1,\ldots,L\right\} $
are independent with identical distribution, the sequence of random
variables $\{Y_{L}^{\left(\Xi\left(0\right)\right)}:L\in\mathbb{N}\}$
with the titled distribution defined in Eq. \eqref{eq:titled} obeys
the central limit theorem. Thus, the results in \cite{Bucklew1993}
can be used to give an accurate estimation of $\psi$ as
\[
\psi=\frac{1}{\sqrt{2\pi L}\sigma\Xi\left(0\right)},
\]
for any $L\in\mathbb{N}$, with
\[
\lim_{L\rightarrow\infty}\frac{1}{L}\ln\frac{p_{\mathrm{out}}}{\psi}=-\Lambda^{*}\left(0\right).
\]
Therefore, we have
\[
\lim_{L\rightarrow\infty}\frac{1}{p_{\mathrm{out}}}\left(\frac{1}{\sqrt{2\pi L}\sigma\Xi\left(0\right)}e^{-L\Lambda^{*}\left(0\right)}\right)=\lim_{L\rightarrow\infty}\frac{p_{\mathrm{ex}}^{\mathrm{upper}}}{p_{\mathrm{out}}}=1.
\]
According to the definition of limit, for any given $\varepsilon>0$,
there is a number $L^{*}\in\mathbb{N}$ such that 
\[
\frac{p_{\mathrm{ex}}^{\mathrm{upper}}}{p_{\mathrm{out}}}-1<\varepsilon,
\]
holds for any $L\geq L^{*}$. Therefore, for any $L\in\mathbb{N}$,
the outage probability $p_{\mathrm{out}}$ is upper bounded by $p_{\mathrm{ex}}^{\mathrm{upper}}$,
and the accuracy of $p_{\mathrm{ex}}^{\mathrm{upper}}$ increases
as $L$ increases for any given $\gamma$.

To sum up, we have 
\begin{equation}
\begin{aligned}p_{\mathrm{out}}\lesssim & p_{\mathrm{ex}}^{\mathrm{upper}}=\frac{1}{\sqrt{2\pi L}\sigma\Xi\left(0\right)}e^{-L\Lambda^{*}\left(0\right)}\\
 & =\frac{1}{\sqrt{2\pi}}e^{-L\left[\Lambda^{*}\left(0\right)+\frac{\ln\left(\sigma\Xi\left(0\right)\right)}{L}+\frac{\ln L}{2L}\right]},
\end{aligned}
\label{eq:pout upper}
\end{equation}
in the sense of $L$ and $\gamma$, where the symbol ``$\lesssim$''
is defined in Eq. \eqref{eq:leq_app}.

\section{Proof of Theorem \ref{thm:lower bound 1}\label{app:lower bound 1}}

Define a random variable $Y$ by letting
\[
Y=\sum_{l=1}^{L}\ln\left(1+\left|h_{l}\right|^{2}\gamma\right),
\]
whose moment generating function is given by
\[
M\left(\xi\right)=\left[e^{\frac{1}{\gamma}}\gamma^{\xi}\Gamma\left(1+\xi,\gamma^{-1}\right)\right]^{L}.
\]
Clearly, we have $p_{\mathrm{out}}=\Pr\left\{ Y<R\right\} $.

Let $F\left(y\right)$ denote the distribution of $Y$, define the
titled distribution for $Y$ as
\[
dF^{\left(\xi\right)}\left(y\right)=\frac{e^{\xi y}dF\left(y\right)}{M\left(\xi\right)}.
\]
For convenience, we let $F_{\alpha}\left(y\right)=F^{\left(\Xi\left(\alpha R\right)\right)}\left(y\right)$,
and use $O_{\alpha}$ to denote the operation $O$ under the titled
distribution $F_{\alpha}\left(y\right)$. Let $Z$ be a random variable
with the distribution of $F_{\alpha}\left(y\right)$, then for any
$\alpha$
\[
\begin{aligned}\mathsf{E}_{\alpha}\left\{ Z\right\}  & =\frac{1}{M\left(\Xi\left(\alpha R\right)\right)}\int_{-\infty}^{+\infty}ye^{\xi y}dF\left(y\right)\\
 & =\left.\frac{\partial}{\partial\xi}\ln M\left(\xi\right)\right|_{\xi=\Xi\left(\alpha R\right)}=\alpha R.
\end{aligned}
\]
Therefore, for any $0<\delta<\alpha<1$, we have
\[
\begin{aligned}F\left(R\right) & =M\left(\Xi\left(\alpha R\right)\right)\int_{0}^{R}e^{-t\Xi\left(\alpha R\right)}dF_{\alpha}\left(t\right)\\
 & \geq M\left(\Xi\left(\alpha R\right)\right)\int_{\delta R}^{R}e^{-t\Xi\left(\alpha R\right)}dF_{\alpha}\left(t\right)\\
 & \geq M\left(\Xi\left(\alpha R\right)\right)e^{-\delta RM\left(\Xi\left(\alpha R\right)\right)}\left[F_{\alpha}\left(R\right)-F_{\alpha}\left(\delta R\right)\right].
\end{aligned}
\]
Since $\delta R<\mathsf{E}_{\alpha}\left\{ Z\right\} <R$, by applying
Cram\'er's theorem \cite{Dembo1998}, we have
\begin{equation}
\left\{ \begin{aligned} & F_{\alpha}\left(\delta R\right)\leq e^{-\Lambda_{\alpha}\left(\delta R\right)};\\
 & F_{\alpha}\left(R\right)\leq1-e^{-\Lambda_{\alpha}\left(R\right)}.
\end{aligned}
\right.\label{eq:chernoff}
\end{equation}
Therefore, we have
\[
\begin{aligned} & F\left(R\right)\geq M\left(\Xi\left(\alpha R\right)\right)e^{-\delta R\Xi\left(\alpha R\right)}\left[F_{\alpha}\left(R\right)-F_{\alpha}\left(\delta R\right)\right]\\
 & =e^{\ln\left(M\left(\Xi\left(\alpha R\right)\right)\right)-\delta R\Xi\left(\alpha R\right)}\left[F_{\alpha}\left(R\right)-F_{\alpha}\left(\delta R\right)\right]\\
 & \geq\left(1-e^{-\Lambda_{\alpha}\left(R\right)}-e^{-\Lambda_{\alpha}\left(\delta R\right)}\right)e^{-\Lambda^{*}\left(\alpha R\right)+\Xi\left(\alpha R\right)R\left(\alpha-\delta\right)}.
\end{aligned}
\]
Clearly, $\Lambda^{*}\left(R\right)$ is given by
\[
\begin{aligned}\Lambda^{*}\left(R\right)= & L\left[-\left(C_{\mathrm{awgn}}-\overline{R}\right)\Xi\left(R\right)+\Xi\left(R\right)\ln\left(1+\frac{1}{\gamma}\right)\right.\\
 & \left.-\ln\left(e^{\frac{1}{\gamma}}\Gamma\left(1+\Xi\left(R\right),\gamma^{-1}\right)\right)\right],
\end{aligned}
\]
and $\Xi\left(R\right)$ is the solution of
\[
R-\frac{L}{\Gamma\left(1+\xi,\gamma^{-1}\right)}G_{2,3}^{3,0}\left(\frac{1}{\gamma}\left|\begin{aligned} & 1,1\\
 & 0,0,1+\xi
\end{aligned}
\right.\right)=0.
\]
Therefore, we have 
\[
\Lambda^{*}\left(\alpha R\right)-\Xi\left(\alpha R\right)R\left(\alpha-\delta\right)=LE_{1}^{\alpha}\left(R,\gamma\right).
\]

In the following, $\Lambda_{\alpha}\left(R\right)$ and $\Lambda_{\alpha}\left(\delta R\right)$
will be obtained from $\Xi\left(R\right)$. Notice that
\[
\frac{\partial}{\partial R}M\left(\Xi\left(R\right)\right)=RM\left(\Xi\left(R\right)\right)\Xi'\left(R\right).
\]
By computing the integration of both sides for the above formula,
we have
\[
\int_{R}^{\mathsf{E}\left(X\right)}\frac{1}{M\left(\Xi\left(t\right)\right)}\frac{\partial}{\partial R}M\left(\Xi\left(t\right)\right)dt=\int_{R}^{\mathsf{E}\left(X\right)}t\Xi'\left(t\right)dt.
\]
Since $\Xi\left(\mathsf{E}\left(X\right)\right)=0$, the integration
equation will become
\[
\ln M\left(\Xi\left(R\right)\right)=R\Xi\left(R\right)+\int_{R}^{\mathsf{E}\left(X\right)}\Xi\left(t\right)dt.
\]
According to the definition of Legendre-Fenchel transform, we have
\begin{equation}
\Lambda^{*}\left(R\right)=-\int_{R}^{\mathsf{E}\left\{ Y\right\} }\Xi\left(t\right)dt.\label{eq:rate function}
\end{equation}
The moment generating function for $Z$ is given by
\[
\mathsf{E}_{\alpha}\left\{ e^{\xi Z}\right\} =\frac{M\left(\xi+\Xi\left(\alpha R\right)\right)}{M\left(\Xi\left(\alpha R\right)\right)}.
\]
Then, $\Xi_{\alpha}\left(s\right)$ satisfies
\[
\begin{aligned}z & =\left.\frac{\partial}{\partial\xi}\frac{M\left(\xi+\Xi\left(\alpha R\right)\right)}{M\left(\Xi\left(\alpha R\right)\right)}\right|_{\text{\ensuremath{\xi}=\ensuremath{\Xi_{\alpha}\left(z\right)}}}\\
 & =\left.\frac{\partial}{\partial\theta}\ln M\left(\theta\right)\right|_{\theta=\Xi_{\alpha}\left(z\right)+\Xi\left(\alpha R\right)}.
\end{aligned}
\]
Therefore, we have
\[
\Xi_{\alpha}\left(z\right)=\Xi\left(z\right)-\Xi\left(\alpha R\right)
\]
According to Eq. \eqref{eq:rate function}, we have
\[
\begin{aligned}\Lambda_{\alpha}^{*}\left(\delta R\right) & =-\int_{\delta R}^{\mathsf{E_{\alpha}\left\{ Z\right\} }}\Xi_{\alpha}\left(t\right)dt\\
 & =R\left(\alpha-\delta\right)\Xi\left(\alpha R\right)-\int_{\delta R}^{\alpha R}\Xi\left(t\right)dt,
\end{aligned}
\]
and $\Lambda_{\alpha}^{*}\left(R\right)=\left.\Lambda_{\alpha}^{*}\left(\delta R\right)\right|_{\delta=1}$. 

Next, we will prove there exists $\alpha$ and $\delta$ such that
$1-e^{-\Lambda_{\alpha}\left(R\right)}-e^{-\Lambda_{\alpha}\left(\delta R\right)}>0$
for any $\overline{R}<\overline{C}$. According to the monotonicity
of $\Xi_{\alpha}\left(R\right)$, we have
\[
\frac{\partial}{\partial\delta}\Lambda_{\alpha}^{*}\left(\delta R\right)=R\left(\Xi\left(\delta R\right)-\Xi\left(\alpha R\right)\right)<0,
\]
and
\[
\frac{\partial^{2}}{\partial\delta^{2}}\Lambda_{\alpha}^{*}\left(\delta R\right)=R^{2}\Xi'\left(\delta R\right)>0.
\]
Therefore, $\Lambda_{\alpha}^{*}\left(\delta R\right)$ achieves the
maximum value at $\delta=0$. According to the result in \cite{Dembo1998},
we have
\[
\inf\left\{ z:z\in\mathbb{R},F_{\alpha}\left(z\right)>0\right\} =0,
\]
and $F_{\alpha}\left(z\right)=e^{-\Lambda_{\alpha}^{*}\left(0\right)}$.
On the other hand,
\[
\lim_{R\rightarrow0^{+}}F_{\alpha}\left(R\right)=\frac{1}{M\left(\Xi\left(\alpha R\right)\right)}\lim_{R\rightarrow0^{+}}\int_{0}^{R}e^{\Xi\left(\alpha R\right)t}dF\left(t\right)=0.
\]
Therefore, we have $\lim_{\delta\rightarrow0}\Lambda_{\alpha}^{*}\left(\delta R\right)=\infty$.
Furthermore,
\[
\frac{\partial}{\partial\alpha}\Lambda_{\alpha}^{*}\left(R\right)=-R^{2}\left(1-\alpha\right)\Xi'\left(\alpha R\right)\leq0.
\]
Consider the result that $\left.\Lambda_{\alpha}^{*}\left(R\right)\right|_{\alpha=1}=0$,
there exists $\epsilon>0$ and $0<\alpha^{*}<1$ such that $\Lambda_{\alpha^{*}}^{*}\left(R\right)=\epsilon$.
Then, $\delta^{*}$ can be chosen as follows:
\[
\Lambda_{\alpha^{*}}^{*}\left(\delta^{*}R\right)>-\ln\left(1-e^{-\epsilon}\right).
\]
Therefore, there is $\alpha^{*}$ and $\delta^{*}$ satisfying
\[
1-e^{-\Lambda_{\alpha^{*}}\left(R\right)}-e^{-\Lambda_{\alpha^{*}}\left(\delta^{*}R\right)}>0.
\]

Finally, we determine the relationship between $\alpha$ and $\delta$.
By letting
\[
\frac{\partial}{\partial\alpha}\left(1-e^{-\Lambda_{\alpha}\left(R\right)}-e^{-\Lambda_{\alpha}\left(\delta R\right)}\right)e^{-\Lambda^{*}\left(\alpha R\right)+\Xi\left(\alpha R\right)R\left(\alpha-\delta\right)}=0,
\]
Eq. \eqref{eq:delta} can then be obtained. As a matter of fact,
the main method of this proof is based on \cite{Theodosopoulos2007}.

\section{Proof of Theorem \ref{thm:Meijer's G bounds}\label{app:Meijer's G bounds}}

Define a function $F_{L}\left(z\right)$ as
\[
F_{L}\left(z\right)=1-\int_{\mathcal{D}\left(z,0\right)}e^{-\sum_{l=1}^{L}x_{l}}d\bm{x},
\]
where the integration domain is defined by
\[
\mathcal{D}\left(z,\eta\right)=\left\{ \bm{x}\left|\prod_{l=1}^{L}x_{l}<z,\bm{x}>\eta\right.\right\} .
\]

Consider the tight upper bound in the high SNR regime, it is easy
to verify that
\[
\begin{aligned}p_{\mathrm{out}} & =\Pr\left\{ \sum_{l=1}^{L}\ln\left(1+\left|h_{l}\right|^{2}\gamma\right)<R\right\} \\
 & \approx\Pr\left\{ \prod_{l=1}^{L}\left|h_{l}\right|^{2}<\frac{e^{R}-1}{\gamma^{L}}\right\} .
\end{aligned}
\]
and
\begin{align*}
p_{\mathrm{out}} & =\Pr\left\{ \sum_{l=1}^{L}\ln\left(1+\left|h_{l}\right|^{2}\gamma\right)<R\right\} \\
 & \leq\Pr\left\{ \ln\left(1+\gamma^{L}\prod_{l=1}^{L}\left|h_{l}\right|^{2}\right)<R\right\} \\
 & =\int_{\mathcal{D}\left(\frac{e^{R}-1}{\gamma^{L}},0\right)}e^{-\sum_{l=1}^{L}x_{l}}d\bm{x}\\
 & =1-F_{L}\left(\frac{e^{R}-1}{\gamma^{L}}\right).
\end{align*}
Therefore, we have
\[
p_{\mathrm{out}}\lesssim1-F_{L}\left(\frac{e^{R}-1}{\gamma^{L}}\right),
\]
holds in the high SNR regime.

Before deriving the tight lower bound in the high SNR regime, we first
rewrite the outage probability as

\begin{align*}
p_{\mathrm{out}} & =\Pr\left\{ \sum_{l=1}^{L}\ln\left(1+\left|h_{l}\right|^{2}\gamma\right)<R\right\} \\
 & =\Pr\left\{ \prod_{l=1}^{L}\left(\frac{1}{\gamma}+\left|h_{l}\right|^{2}\right)<\frac{e^{R}}{\gamma^{L}}\right\} \\
 & =e^{\frac{L}{\gamma}}\int_{\substack{\mathcal{D}}
\left(\frac{e^{R}}{\gamma^{L}},\frac{1}{\gamma}\right)}e^{-\sum_{l=1}^{L}x_{l}}d\bm{x}.
\end{align*}
Next, the principle of mathematical induction will be used to show
that $p_{\mathrm{out}}\geq1-e^{\frac{L}{\gamma}}F_{L}\left(e^{R}\gamma^{-L}\right)$.
For $L=1$, it is easy to verify that 
\[
p_{\mathrm{out}}=1-e^{-\frac{e^{R}-1}{\gamma}}=1-e^{\frac{1}{\gamma}}F_{1}\left(\frac{e^{R}}{\gamma}\right).
\]
Assume the proposition holds for $L=k,\, k\geq1$. Then, for the situation
of $L=k+1$, the lower bound of $p_{\mathrm{out}}$ is calculated
as follows: 
\[
\begin{aligned}p_{\mathrm{out}} & =e^{\frac{k+1}{\gamma}}\int_{\substack{\mathcal{D}}
\left(\frac{e^{R}}{\gamma^{k+1}},\frac{1}{\gamma}\right)}e^{-\sum_{l=1}^{k+1}x_{l}}d\bm{x}\\
 & =e^{\frac{1}{\gamma}}\int_{\frac{1}{\gamma}}^{\infty}e^{-x_{1}}\left[\int_{\substack{\mathcal{D}}
\left(\frac{e^{R}}{\gamma^{k+1}x_{1}},\frac{1}{\gamma}\right)}e^{-\sum_{l=1}^{k}x_{l}}d\bm{x}'\right]dx_{1}\\
 & \geq e^{\frac{1}{\gamma}}\int_{\frac{1}{\gamma}}^{\infty}e^{-x_{1}}\left[1-e^{\frac{k}{\gamma}}F_{k}\left(\frac{e^{R}}{\gamma^{k+1}x_{1}}\right)\right]dx_{1}\\
 & =1-e^{\frac{k+1}{\gamma}}\int_{0}^{\infty}e^{-x_{1}}F_{k}\left(\frac{e^{R}}{\gamma^{k+1}x_{1}}\right)dx_{1}\\
 & =1-e^{\frac{k+1}{\gamma}}F_{k+1}\left(\frac{e^{R}}{\gamma^{k+1}}\right),
\end{aligned}
\]
where $\bm{x}'=\left[x_{2},\ldots,x_{k+1}\right]$. According to the
principle of mathematical induction, the inequality
\[
p_{\mathrm{out}}\gtrsim1-e^{\frac{L}{\gamma}}F_{L}\left(\frac{e^{R}}{\gamma^{L}}\right)
\]
holds for any $L\in\mathbb{N}$.

Next, we will show that $F_{L}\left(z\right)$ is the Meijer's $G$-function
as shown in the theorem. The proof is also based on the principle
of mathematical induction. For $L=1$,

\[
F_{1}\left(z\right)=1-\int_{0}^{z}e^{-x}dx=e^{-z}=G_{0,1}^{1,0}\left(z\left|\begin{array}{c}
-\\
0
\end{array}\right.\right),
\]
where the last equation follows that 
\[
\begin{aligned}G_{0,1}^{1,0}\left(z\left|\begin{aligned}-\\
0
\end{aligned}
\right.\right) & =\frac{1}{2\pi i}\oint_{\mathcal{L}}\Gamma\left(-s\right)z^{s}ds\\
 & =\sum_{n=0}^{\infty}\frac{\left(-z\right)^{n}}{n!}=e^{-z}.
\end{aligned}
\]
For $L=2$, we have
\[
\begin{aligned}F_{2}\left(z\right) & =1-\iint_{\mathcal{D}\left(z,0\right)}e^{-x_{1}-x_{2}}dx_{1}dx_{2}\\
 & =1-\int_{0}^{\infty}e^{-x_{1}}\left(\int_{0}^{\frac{z}{x_{1}}}e^{-x_{2}}dx_{2}\right)dx_{1}\\
 & =1-\int_{0}^{\infty}e^{-x_{1}}\left(1-e^{-\frac{z}{x_{1}}}\right)dx_{1}\\
 & =\int_{0}^{\infty}e^{-x_{1}-\frac{z}{x_{1}}}dx_{1}\\
 & =2\sqrt{z}K_{1}\left(2\sqrt{z}\right)=G_{0,2}^{2,0}\left(z\left|\begin{array}{c}
-\\
0,1
\end{array}\right.\right),
\end{aligned}
\]
where $K_{\nu}\left(z\right)$ is the modified Bessel function of
the second kind, and the last equation follows Eq. (3) in Chapter
9.34 of \cite{Gradshteyn2007}
\[
G_{0,2}^{2,0}\left(\frac{z^{2}}{4}\left|\begin{aligned}-\\
\frac{\mu-\nu}{2},\frac{\mu+\nu}{2}
\end{aligned}
\right.\right)=2\left(\frac{z}{2}\right)^{\mu}K_{\nu}\left(z\right).
\]
Assume the proposition holds for $L=k,\, k\geq1$. Then, for $L=k+1$,
$F_{k+1}\left(z\right)$ is calculated as 
\[
\begin{aligned}F_{k+1}\left(z\right) & =1-\int_{\mathcal{D}\left(z,0\right)}e^{-\sum_{l=1}^{k+1}x_{l}}d\bm{x}\\
 & =1-\int_{0}^{\infty}e^{-x_{1}}\left[\int_{\mathcal{D}\left(\frac{z}{x_{1}},0\right)}e^{-\sum_{l=1}^{k}x_{l}}d\bm{x}'\right]dx_{1}\\
 & =\int_{0}^{\infty}e^{-x_{1}}G_{0,k}^{k,0}\left(\frac{z}{x_{1}}\left|\begin{array}{c}
-\\
0,1,\ldots,1
\end{array}\right.\right)dx_{1}\\
 & =\frac{1}{2\pi i}\oint_{\mathcal{L}}\Gamma\left(-s\right)\Gamma^{k}\left(1-s\right)z^{s}ds\\
 & =G_{0,k+1}^{k+1,0}\left(z\left|\begin{array}{c}
-\\
0,1,\ldots,1
\end{array}\right.\right).
\end{aligned}
\]
It should be noted that the path $\mathcal{L}$ runs from $-\infty$
to $+\infty$ in such a way that the poles of the functions $\Gamma\left(-s\right)$
and $\Gamma\left(1-s\right)$ lie to the right of $\mathcal{L}$.
Therefore, the path $\mathcal{L}$ can be chosen as $\Re\left(\mathcal{L}\right)=-\frac{1}{2}$,
then $\Re\left(1-s\right)=\frac{3}{2}$ for any $s$ along this path.
Under this condition, $\Gamma\left(1-s\right)$ can be expanded as
$\int_{0}^{\infty}x^{-s}e^{-x}dx$. According to the principle of
mathematical induction, the equation
\[
F_{L}\left(z\right)=G_{0,L}^{L,0}\left(z\left|\begin{array}{c}
-\\
0,1,\ldots,1
\end{array}\right.\right)
\]
holds for any $L\in\mathbb{N}$. As a matter of fact, this proof can
also be carried out by applying the Mellin's transform to the product
of independent random variables \cite{Mathai2010}.

In the low SNR regime, let $R=rC_{\mathrm{awgn}}$, the approximation
is given by
\begin{align*}
p_{\mathrm{out}} & =\Pr\left\{ \sum_{l=1}^{L}\ln\left(1+\left|h_{l}\right|^{2}\gamma\right)<R\right\} \\
 & \approx\Pr\left\{ \ln\left(1+\gamma\sum_{l=1}^{L}\left|h_{l}\right|^{2}\right)<r\ln\left(1+\gamma\right)\right\} \\
 & =\Pr\left\{ \sum_{l=1}^{L}\left|h_{l}\right|^{2}<\frac{\left(1+\gamma\right)^{r}-1}{\gamma}\right\} \\
 & =1-\frac{\Gamma\left(L,r\right)}{\left(L-1\right)!},
\end{align*}
where the property that the sum of exponential distributed random
varibles is an Erlang distributed random varible has been used.

\section{Proof of Theorem \ref{thm:Meijer exponent}\label{app:Meijer exponent}}

Define a function $f\left(\gamma\right)$ as follows:
\[
f\left(\gamma\right)=\ln\left(1-e^{\frac{\lambda}{\gamma}}G_{0,L}^{L,0}\left(\frac{\left(1+\gamma\right)^{r}}{\gamma^{L}}\left|\begin{array}{c}
-\\
0,1,\ldots,1
\end{array}\right.\right)\right).
\]
According to Theorem \ref{thm:Meijer's G bounds} and Proposition
\ref{pro:Meijer's G approximation}, we have
\[
p_{\mathrm{out}}\approx e^{f\left(\gamma\right)}.
\]
For any $x\in\left(0,1\right)$, it is easy to verify that
\[
\left\{ \begin{aligned} & \frac{\partial}{\partial x}\ln\left(1-x\right)=\frac{-1}{1-x}<0;\\
 & \frac{\partial^{2}}{\partial x^{2}}\ln\left(1-x\right)=\frac{-1}{\left(1-x\right)^{2}}<0.
\end{aligned}
\right.
\]
According to Lemma \ref{lem:convex}, the outage probability can be
bounded by
\[
p_{\mathrm{out}}\approx e^{f\left(\gamma\right)}\geq e^{\gamma\frac{\partial}{\partial\gamma}f\left(\gamma\right)}.
\]
According to the definition of Meijer's $G$-function, we have
\[
\begin{aligned} & \frac{\partial}{\partial z}G_{0,L}^{L,0}\left(z\left|\begin{array}{c}
-\\
0,1,\ldots,1
\end{array}\right.\right)\\
= & \frac{\partial}{\partial z}\frac{1}{2\pi i}\oint_{\mathcal{L}}\Gamma\left(-s\right)\Gamma^{L-1}\left(1-s\right)z^{s}ds\\
= & \frac{1}{2\pi i}\oint_{\mathcal{L}}\Gamma\left(-s\right)\Gamma^{L-1}\left(1-s\right)sz^{s-1}ds\\
= & -\frac{1}{2\pi i}\oint_{\mathcal{L}}\left(-1-t\right)\Gamma\left(-1-t\right)\Gamma^{L-1}\left(-t\right)z^{t}dt\\
= & -\frac{1}{2\pi i}\oint_{\mathcal{L}}\Gamma^{L}\left(-t\right)z^{t}dt\\
= & -G_{0,L}^{L,0}\left(z\left|\begin{array}{c}
-\\
0,0,\ldots,0
\end{array}\right.\right).
\end{aligned}
\]
Then, for $\gamma\frac{\partial}{\partial\gamma}f\left(\gamma\right)$
we have 

\[
\begin{aligned}\gamma\frac{\partial}{\partial\gamma}f\left(\gamma\right)= & -L\left[\left(1-\frac{r}{L}\right)+\frac{1}{\gamma}\right]\frac{\left(1+\gamma\right)^{r-1}}{\gamma^{L-1}}\cdot\\
 & \frac{e^{\frac{\lambda}{\gamma}}G_{0,L}^{L,0}\left(\frac{\left(1+\gamma\right)^{r}}{\gamma^{L}}\left|\begin{array}{c}
-\\
0,0,\ldots,0
\end{array}\right.\right)}{1-e^{\frac{\lambda}{\gamma}}G_{0,L}^{L,0}\left(\frac{\left(1+\gamma\right)^{r}}{\gamma^{L}}\left|\begin{array}{c}
-\\
0,1,\ldots,1
\end{array}\right.\right)}+\\
 & \frac{\frac{\lambda}{\gamma}e^{\frac{\lambda}{\gamma}}G_{0,L}^{L,0}\left(\frac{\left(1+\gamma\right)^{r}}{\gamma^{L}}\left|\begin{array}{c}
-\\
0,1,\ldots,1
\end{array}\right.\right)}{1-e^{\frac{\lambda}{\gamma}}G_{0,L}^{L,0}\left(\frac{\left(1+\gamma\right)^{r}}{\gamma^{L}}\left|\begin{array}{c}
-\\
0,1,\ldots,1
\end{array}\right.\right)}.
\end{aligned}
\]
Therefore, Eq. \eqref{eq:Meijer exponent} holds.

\section{Proof of Theorem \ref{cor:dmt}\label{app:dmt}}

From the proof of Theorem \ref{thm:Meijer's G bounds}, we have
\[
\lim_{\gamma\rightarrow\infty}\frac{\frac{\lambda}{\gamma}e^{\frac{\lambda}{\gamma}}G_{0,L}^{L,0}\left(\frac{\left(1+\gamma\right)^{r}}{\gamma^{L}}\left|\begin{array}{c}
-\\
0,1,\ldots,1
\end{array}\right.\right)}{1-e^{\frac{\lambda}{\gamma}}G_{0,L}^{L,0}\left(\frac{\left(1+\gamma\right)^{r}}{\gamma^{L}}\left|\begin{array}{c}
-\\
0,1,\ldots,1
\end{array}\right.\right)}=0.
\]
Since $e^{\frac{\lambda}{\gamma}}\rightarrow1$ and $\frac{1+\gamma}{\gamma}\rightarrow1$
as $\gamma\rightarrow\infty$, according to the rule of L'Hospital,
we have
\[
\begin{aligned} & \lim_{\gamma\rightarrow\infty}d_{\mathrm{f}}^{*}\left(r,\gamma\right)\\
= & d^{*}\left(r\right)\lim_{z\rightarrow0}\frac{zG_{0,L}^{L,0}\left(z\left|\begin{array}{c}
-\\
0,0,\ldots,0
\end{array}\right.\right)}{1-G_{0,L}^{L,0}\left(z\left|\begin{array}{c}
-\\
0,1,\ldots,1
\end{array}\right.\right)}\\
= & d^{*}\left(r\right)\lim_{z\rightarrow0}\left[1+\frac{z\frac{\partial}{\partial z}G_{0,L}^{L,0}\left(z\left|\begin{array}{c}
-\\
0,0,\ldots,0
\end{array}\right.\right)}{G_{0,L}^{L,0}\left(z\left|\begin{array}{c}
-\\
0,0,\ldots,0
\end{array}\right.\right)}\right]\\
= & d^{*}\left(r\right)+d^{*}\left(r\right)\lim_{z\rightarrow0}\frac{G_{0,L}^{L,0}\left(z\left|\begin{array}{c}
-\\
0,0,\ldots,1
\end{array}\right.\right)}{G_{0,L}^{L,0}\left(z\left|\begin{array}{c}
-\\
0,0,\ldots,0
\end{array}\right.\right)},
\end{aligned}
\]
where the last equality follows that
\[
\begin{aligned} & z\frac{\partial}{\partial z}G_{0,L}^{L,0}\left(z\left|\begin{array}{c}
-\\
0,0,\ldots,0
\end{array}\right.\right)\\
= & z\frac{1}{2\pi i}\oint_{\mathcal{L}}\Gamma^{L}\left(-s\right)sz^{s-1}ds\\
= & \frac{1}{2\pi i}\oint_{\mathcal{L}}\Gamma^{L-1}\left(-s\right)\Gamma\left(1-s\right)z^{s}ds\\
= & G_{0,L}^{L,0}\left(z\left|\begin{array}{c}
-\\
0,0,\ldots,1
\end{array}\right.\right).
\end{aligned}
\]
By repeating this process, we have
\[
\begin{aligned} & \lim_{z\rightarrow0}\frac{G_{0,L}^{L,0}\left(z\left|\begin{array}{c}
-\\
0,0,\ldots,1
\end{array}\right.\right)}{G_{0,L}^{L,0}\left(z\left|\begin{array}{c}
-\\
0,0,\ldots,0
\end{array}\right.\right)}\\
= & \lim_{z\rightarrow0}\frac{G_{0,L}^{L,0}\left(z\left|\begin{array}{c}
-\\
1,1,\ldots,1
\end{array}\right.\right)}{G_{0,L}^{L,0}\left(z\left|\begin{array}{c}
-\\
0,1,\ldots,1
\end{array}\right.\right)}.
\end{aligned}
\]

In the following, the principle of mathematical induction will be
used to show the above limit is zero. For $L=1$, we have
\[
\begin{aligned}\lim_{z\rightarrow0}\frac{G_{0,1}^{1,0}\left(z\left|\begin{array}{c}
-\\
1
\end{array}\right.\right)}{G_{0,1}^{1,0}\left(z\left|\begin{array}{c}
-\\
0
\end{array}\right.\right)} & =\lim_{z\rightarrow0}\frac{\frac{1}{2\pi i}\oint_{\mathcal{L}}\Gamma\left(1-s\right)z^{s}ds}{e^{-z}}\\
 & =\frac{\lim_{z\rightarrow0}z\sum_{n=0}^{\infty}\frac{\left(-z\right)^{n}}{n!}}{\lim_{z\rightarrow0}e^{-z}}\\
 & =\frac{0}{1}=0.
\end{aligned}
\]
In the above derivation, we used the property that the poles of $\Gamma\left(1-s\right)$
are $s=n\in\mathbb{N}$, and the corresponding residuals are $\frac{\left(-1\right)^{n}}{n!}$.
Suppose the proposition holds for $L=k$, when $L=k+1$ we have
\[
\begin{aligned} & \lim_{z\rightarrow0}\frac{G_{0,k+1}^{k+1,0}\left(z\left|\begin{array}{c}
-\\
1,1,\ldots,1
\end{array}\right.\right)}{G_{0,k+1}^{k+1,0}\left(z\left|\begin{array}{c}
-\\
0,1,\ldots,1
\end{array}\right.\right)}\\
& = \lim_{z\rightarrow0}\frac{\frac{1}{2\pi i}\oint_{\mathcal{L}}\Gamma^{k+1}\left(1-s\right)z^{s}ds}{\frac{1}{2\pi i}\oint_{\mathcal{L}}\Gamma\left(-s\right)\Gamma^{k}\left(1-s\right)z^{s}ds}\\
& = \lim_{z\rightarrow0}\frac{\frac{1}{2\pi i}\oint_{\mathcal{L}}\Gamma^{k}\left(1-s\right)z^{s}\int_{0}^{\infty}t^{-s}e^{-t}dtds}{\frac{1}{2\pi i}\oint_{\mathcal{L}}\Gamma\left(-s\right)\Gamma^{k-1}\left(1-s\right)z^{s}\int_{0}^{\infty}t^{-s}e^{-t}dtds}\\
& = \lim_{z\rightarrow0}\frac{\int_{0}^{\infty}e^{-t}\frac{1}{2\pi i}\oint_{\mathcal{L}}\Gamma^{k}\left(1-s\right)\left(\frac{z}{t}\right)^{s}dsdt}{\int_{0}^{\infty}e^{-t}\frac{1}{2\pi i}\oint_{\mathcal{L}}\Gamma\left(-s\right)\Gamma^{k-1}\left(1-s\right)\left(\frac{z}{t}\right)^{s}dsdt}\\
& = \frac{\int_{0}^{\infty}e^{-t}\lim_{z\rightarrow0}G_{0,k}^{k,0}\left(\frac{z}{t}\left|\begin{array}{c}
-\\
1,1,\ldots,1
\end{array}\right.\right)dt}{\int_{0}^{\infty}e^{-t}\lim_{z\rightarrow0}G_{0,k}^{k,0}\left(\frac{z}{t}\left|\begin{array}{c}
-\\
0,1,\ldots,1
\end{array}\right.\right)dt}=\frac{0}{1}=0.
\end{aligned}
\]
The integration path $\mathcal{L}$ is $\Re\left(\mathcal{L}\right)=-\frac{1}{2}$.
According to the principle of mathematical induction, the limit is
zero for any $L\in\mathbb{N}$.

\bibliographystyle{IEEEtran}
\bibliography{library}

\vfill
\begin{IEEEbiographynophoto}
{Bo Bai} (S'09-M'11) received his BS degrees in Department of Communication Engineering with the highest honor from Xidian University in 2004, Xi'an China. He also obtained the honor of \emph{Outstanding Graduates of Shaanxi Province}. He received the Ph.D degree in Department of Electronic Engineering from Tsinghua University in 2010, Beijing China. He also obtained the honor of \emph{Young Academic Talent of Electronic Engineering} in Tsinghua University. From 2009 to 2012, he was a visiting research staff (Research Assistant from April 2009 to September 2010 and Research Associate from October 2010 to April 2012) in Department of Electronic and Computer Engineering, Hong Kong University of Science and Technology (HKUST). Now he is an Assistant Professor in Department of Electronic Engineering, Tsinghua University. He has also obtained the support from \emph{Backbone Talents Supporting Project} of Tsinghua University.

His research interests include hot topics in wireless communications, information theory, random graph, and combinatorial design. He has served as a TPC member for IEEE ICC 2010, IEEE ICC 2012, IEEE ICCC 2012, and IEEE ICCVE 2012, and also a Session Chair for IEEE Globecom '08. He has also served as a reviewer for a number of major IEEE journals and conferences. He received \emph{Student Travel Grant} at IEEE Globecom '09. He was also invited as \emph{Young Scientist Speaker} at IEEE TTM 2011.
\end{IEEEbiographynophoto}

\begin{IEEEbiographynophoto}
{Wei Chen} (S'03-M'07) received his BS and PhD degrees in Electronic Engineering (both with the highest honors) from Tsinghua University, Beijing, China, in 2002, and 2007, respectively. From 2005 to 2007, he was also a visiting research staff in the Hong Kong University of Science and Technology (HKUST). Since July 2007, he has been with Department of Electronic Engineering, Tsinghua University, where he is currently an Associate Professor and the vice director of institute of communications. He has ever visited Southampton University, HKUST, and the Chinese University of Hong Kong.

His research interests are in broad areas of wireless communications, information theory and applied optimizations. He served as an Editor for IEEE Wireless Communications Letters, an vice director of youth committee of China institute of communications, a tutorial Co-chair of the 2013 IEEE International Conference on Communications, a track Co-chair of the wireless track in the 2013 IEEE CCNC, a TPC Co-chair of the 2011 Spring IEEE Vehicular Technology Conference, the Publication Chair of the 2012 IEEE International Conference on Communications in China (ICCC), a TPC Co-chair of the Wireless Communication Symposium at the 2010 IEEE International Conference on Communications (ICC), a Student Travel Grant Chair of ICC 2008. He is the chief scientist of a national 973 young scientist project. He was the recipient of the 2010 IEEE Comsoc Asia Paciﬁc Board Best Young Researcher Award, the 2009 IEEE Marconi Prize Paper Award, the Best Paper Award at IEEE ICC 2006, the Best Paper Award at the 2007 IEEE IWCLD, the 2011 Tsinghua Raising Academic Star Award, the 2012 Tsinghua Young Faculty Teaching Excellence Award, the First Prize in the first national young faculty teaching competition, the First Prize in the Seventh Beijing Young Faculty Teaching Competition, and the First Prize in the Fifth Tsinghua University Young Faculty Teaching Competition.
\end{IEEEbiographynophoto}

\begin{IEEEbiographynophoto}
{Khaled B. Letaief} (S'85-M'86-SM'97-F'03) received the BS degree with distinction in Electrical Engineering from Purdue University at West Lafayette, Indiana, USA, in December 1984.  He received the MS and Ph.D. Degrees in Electrical Engineering from Purdue University, in August 1986, and May 1990, respectively.  From January 1985 and as a Graduate Instructor in the School of Electrical Engineering at Purdue University, he has taught courses in communications and electronics.

From 1990 to 1993, he was a faculty member at the University of Melbourne, Australia.  Since 1993, he has been with the Hong Kong University of Science \& Technology (HKUST) where he is currently the Dean of Engineering.  He is also Chair Professor of Electronic and Computer Engineering as well as the Director of the Hong Kong Telecom Institute of Information Technology.  His current research interests include wireless and mobile networks, Broadband wireless access, Cooperative networks, Cognitive radio, and Beyond 3G systems.  In these areas, he has over 400 journal and conference papers and given invited keynote talks as well as courses all over the world. He has also 3 granted patents and 10 pending US patents.

Dr. Letaief served as consultants for different organizations and is the founding Editor-in-Chief of the \emph{IEEE Transactions on Wireless Communications}.  He has served on the editorial board of other prestigious journals including the \emph{IEEE Journal on Selected Areas in Communications – Wireless Series} (as Editor-in-Chief).  He has been involved in organizing a number of major international conferences and events. These include serving as the Co-Technical Program Chair of the \emph{2004 IEEE International Conference on Communications, Circuits and Systems}, ICCCS’04; General Co-Chair of the 2007 \emph{IEEE Wireless Communications and Networking Conference}, WCNC’07; Technical Program Co-Chair of the 2008 \emph{IEEE International Conference on Communication}, ICC’08, Vice General Chair of the 2010 \emph{IEEE International Conference on Communication}, ICC’10, and General Co-Chair of 2011 \emph{IEEE Technology Time machine}, TTM’11.

He served as an elected member of the IEEE Communications Society Board of Governors, IEEE Distinguished lecturer, and Vice-President for Conferences of the IEEE Communications Society.  He also served as the Chair of the IEEE Communications Society Technical Committee on Wireless Communications, Chair of the Steering Committee of the \emph{IEEE Transactions on Wireless Communications}, and Chair of the 2008 IEEE Technical Activities/Member and Geographic Activities Visits Program.  He served as member of the IEEE Communications Society and IEEE Vehicular Technology Society Fellow Evaluation Committees as well as member of the IEEE Technical Activities Board/PSPB Products \& Services Committee.

He is the recipient of many distinguished awards including the Michael G. Gale Medal for Distinguished Teaching (Highest university-wide teaching award at HKUST); 2007 IEEE Communications Society Publications Exemplary Award, 2009 IEEE Marconi Prize Award in Wireless Communications, 2010 Outstanding Electrical and Computer Engineer Award by Purdue University, 2011 IEEE Communications Society Harold Sobol Award, 2011 IEEE Communications Society Wireless Communications Committee Recognition Award, and 9 IEEE Best Paper Awards.

Dr. Letaief is a \emph{Fellow} of IEEE and is currently serving as Member of the IEEE Product Services and Publications Board, and the Treasurer of the IEEE Communications Society. He is also recognized by Thomson Reuters as an \emph{ISI Highly Cited Researcher}.
\end{IEEEbiographynophoto}

\begin{IEEEbiographynophoto}
{Zhigang Cao} (M'84-SM'85) graduated with golden medal from the Department of Radio Electronics at Tsinghua University, Beijing in 1962. Since then he has been with Tsinghua University, where he is currently a professor of Electronic Engineering Department. He was a visiting scholar at Stanford University from 1984 to 1986, and visiting professor at Hong Kong University of Science and Technology in 1997.

He has published six books and more than 500 papers on Communications and Signal Processing fields, and held over 20 patents. He has won 11 research awards and special grant from Chinese government for his outstanding contributions to education and research. He is a co-recipient of several best paper awards including the 2009 IEEE Marconi Prize Paper Award. His current research interests include mobile communications and satellite communications.

Prof. Cao is a CIC fellow, IEEE senior member, CIE senior member and IEICE member. He currently serves as an associate editor-in-chief of ACTA Electronics Sinica. He also serves as the editor of China Communications, Journal of Astronautics, and Frontiers of Electrical and Electronic Engineering.
\end{IEEEbiographynophoto}
\end{document}